\documentclass[journal]{IEEEtran}
\usepackage{authblk}
\normalsize
\usepackage{graphicx}
\usepackage{subcaption}
\usepackage{hyperref}
\usepackage{url}
\usepackage{array}
\usepackage[utf8]{inputenc} 
\usepackage[T1]{fontenc}
\usepackage{amsmath}
\usepackage{mathtools}
\usepackage{blkarray}
\usepackage{amsfonts}
\usepackage{amssymb}
\usepackage{amsthm}
\usepackage{enumerate}
\usepackage{cite}
\usepackage{lipsum}
\usepackage{mathtools}
\usepackage{cuted}
\usepackage{calc}
\usepackage{color}
\usepackage{epsfig}
\usepackage{setspace}
\usepackage{pstricks}
\usepackage{cancel}
\usepackage{multirow}
\usepackage{mathrsfs}
\usepackage{algorithm}
\usepackage{algpseudocode}
\usepackage{multirow}
\usepackage{diagbox}
\usepackage{multicol}
\usepackage{booktabs,makecell}
\usepackage{mathrsfs}
\usepackage{enumitem}
\newtheorem{defn}{Definition}
\newtheorem{thm}{{\cal T}heorem}
\newtheorem{cor}{Corollary}

\newtheorem{property}{Property}

\newtheorem{remark}{Remark}

\newtheorem{example}{Example}

\newcommand{\U}{\mathcal{U}}

\newcommand{\R}{\mathcal{R}}
\newcommand{\W}{\mathcal{W}}
\newcommand{\s}{\mathcal{S}}

\newcommand{\F}{\mathcal{F}}
\newcommand{\M}{\mathcal{M}}

\newcommand{\Q}{\mathcal{Q}}
\newcommand{\N}{\mathcal{N}}
\newcommand{\Y}{\mathcal{Y}}

\input{epsf.sty}

\title{Multi-access Distributed Computing Models from Map-Reduce Arrays}
\begin{document}
	\author{
	\IEEEauthorblockN{Shanuja Sasi, ~\IEEEmembership{Member, ~IEEE},
		Onur Günlü, ~\IEEEmembership{Member, ~IEEE} 
		and B. Sundar Rajan, ~\IEEEmembership{Fellow, ~IEEE}}
    \thanks{S. Sasi and O. Günlü are with the Information Theory and Security Laboratory, Division of Information Coding, Linköping University, SE-581 83 Linköping, Sweden (e-mail: \{shanuja.sasi, onur.gunlu\}@liu.se).}
    \thanks{ B. S. Rajan is with the Department of Electrical Communication Engineering, Indian Institute of Science, Bangalore, 560012, India (e-mail: bsrajan@iisc.ac.in).}
    }
	\maketitle
	\thispagestyle{empty}	
\begin{abstract}
	A novel distributed computing model called \textit{Multi-access Distributed Computing (MADC)} was recently introduced in [B. Federico and P. Elia, ``Multi-Access Distributed Computing,''  June 2022, [online] Available: http://www.arXiv:2206.12851]. The MADC models with Combinatorial Topology (CT) was studied, where there are $\Lambda$ mapper nodes and $K = {\Lambda \choose \alpha}$ reducer nodes with each reducer node connected  to distinct $\alpha$ mapper nodes. In this paper, we represent MADC models via 2-layered bipartite graphs called Map-Reduce Graphs (MRGs) and a  set of arrays called Map-Reduce Arrays (MRAs) inspired from the Placement Delivery Arrays (PDAs) used in the coded caching literature. The connection between  MRAs and MRGs is established, thereby exploring new topologies and providing coded shuffling schemes for the MADC models with MRGs  using the structure of MRAs. A novel \textit{Nearest Neighbor Connect-MRG (NNC-MRG)} is explored and a coding scheme is provided for MADC models with NNC-MRG, exploiting the connections between MRAs and PDAs. Moreover, CT is generalized to Generalized Combinatorial-MRG (GC-MRG). A set of $g-$regular MRAs is provided which corresponds to the existing scheme for MADC models with CT and extended those to generate another set of MRAs to represent  MADC models with GC-MRG. A lower bound on the computation-communication curve for MADC model with GC-MRG under homogeneous setting is derived and certain cases are explored where the existing scheme is optimal under CT. One of the major limitations of the existing scheme for CT is that it requires an exponentially large number of reducer nodes and input files for large $\Lambda$. This can be overcome by representing CT by MRAs, where  coding schemes can be derived even if some of the reducer nodes are not present. Another way of tackling this is by using a different MRG, specifically NNC-MRG, where the number of reducer nodes and files required are significantly smaller compared to CT. Hence, the advantages  are two-fold, which is achievable at the expense of a slight increase in the communication load. 
\end{abstract}
\begin{IEEEkeywords}
	Distributed Computing, Map-Reduce Framework, Placement Delivery Array.
\end{IEEEkeywords}
\section{introduction}
\label{intro}
 The commonly-used Distributed Computing (DC) frameworks, such as Hadoop Map-Reduce \cite{Mapreduce} (widely used by Google, Facebook, Amazon etc.) and Apache Spark \cite{apache}, divide the computing tasks into multiple parallel tasks and distribute them across the servers. When distributing a set of functions across the servers, computation cost and communication cost among servers, become important metrics. A common computing framework, called Map-Reduce \cite{Mapreduce} framework, deals with computation tasks that involve large data sizes. In such a framework, we have a set of servers and the computing task is carried out  by these servers in three stages, i.e., Map, Shuffle, and Reduce stages. Initially, each input data block (file) is stored multiple times across the servers, and each server processes the locally stored data to generate some intermediate values (IV) in the Map stage. In the Shuffle stage, servers exchange the IVs among themselves so that the final output functions are calculated distributedly across the servers in the Reduce stage. The output functions to be computed are assumed to be a function of the input data blocks (files).

Coding-theoretic techniques have been widely used in DC, for numerous applications including distributed storage \cite{DGWWR}, caching \cite{MAN}, coded matrix multiplication \cite{LMADC,MCJ}, and gradient computations \cite{OGU,OUG,WGLL,TLDK,SAR}. However, most of these approaches are application-specific. In \cite{LMA},  the authors used a methodology, called Coded Distributed Computing (CDC),  to exploit coding in data shuffling, which can be applied to any DC framework that has a Map-Reduce structure. The CDC helps to reduce the communication load as compared to uncoded schemes by a factor of the computation load in Map-Reduce framework. For the CDC, the files are stored multiple times across the servers in the Map phase to enable coding opportunities in the Shuffling phase. In  \cite{YYW},  the authors used placement delivery array (PDA) designs to construct a coded computing scheme and they characterized the storage-computation-communication trade-off, rather than the computation-communication trade-off in \cite{LMA}. The CDC has been extensively studied in the literature \cite{SFZ,YGK,YL,YSD,LCW,WCJ,DZZWL,LMAFog,PLSSM,SZZG,WCJnew}, some of which are based on PDAs\cite{JQ,YTC,RK}. PDAs are originally introduced in \cite{YCTCPDA} as a solution for coded caching problems, but it is now a widely used design for various problems. 

In \cite{BP}, a new model was studied, called Multi-access Distributed Computing (MADC) model (as shown as in Fig. \ref{fig: madc model}), where there are two sets of nodes, i.e., mapper and reducer nodes. Unlike in the original setting \cite{LMA} where mapper  and reducer nodes are the same, in \cite{BP}, mapper and reducer nodes are two different entities and each reducer node is connected to multiple mapper nodes. During the Map phase, files are stored across the mapper nodes that compute the IVs. Reducer nodes collect the IVs from the mapper nodes to which they are connected, exchange IVs among themselves, and then calculate the output functions. In \cite{BP}, the authors considered the setting where reducer nodes are connected to mapper nodes  with Combinatorial Topology (CT).  Each reducer node is connected uniquely to $\alpha$ mapper nodes, i.e., there is exactly one reducer for each set of $\alpha$ mapper nodes.

 The main contributions of this paper can be summarized as follows. 
 \begin{itemize}
 	\item We define a new 2-layered bipartite graph and a new array named as Map-Reduce Graph (MRG) and Map-Reduce Array (MRA), respectively, to represent MADC models.
 	\item We connect MRAs to MRGs, apply the MRA design to represent the shuffle and reduce phases for the corresponding MADC models and provide a new coding scheme with the help of the MRA structure, thereby discovering new topologies. 
 	\item We define a new set of MRGs named  Nearest Neighbor Connect-MRGs (NNC-MRGs), prove that a set of $l-$cyclic $g-$regular PDAs represents these MRGs and provide coding scheme for MADC models with NNC-MRGs.
 	\item We consider a generalized version of CT named as Generalized Combinatorial-MRGs (GC-MRGs). We prove that a set of $g-$regular MRAs corresponds to the existing scheme for MADC models with CT that achieves the computation-communication corner points given in (\ref{BE rate}) and further extend those  $g-$regular MRAs to generate another set of MRAs to represent  MADC models with GC-MRGs.
 	\item In \cite{BP}, a lower bound on the communication load was derived for general heterogeneous networks, where nodes have varying storage and computing capabilities. We assume the network to be homogeneous, i.e., the computing nodes (mapper nodes) of the network have the same amount of storage and computation resources and obtain a lower bound for MADC models with GC-MRGs. 
 	\item One of the major advantages of representing CT by MRAs is that  we can have coding schemes for MADC models with CT even if some of the reducer nodes are not present. 
 	\item We also observe  that the NNC-MRG is better than the CT in terms of flexibility in choosing the values of number of reducer nodes and files.
 \end{itemize}
 {\it Organization of this paper:} We define the problem under consideration in Section \ref{problem defintion}. In Section \ref{MRA}, we define MRGs, MRAs and establish the relation between MRAs and PDAs. We connect MRAs to MRGs in Section \ref{NT}. The NNC-MRG is defined in section \ref{NNC-MRG} and a set of $l-$cyclic $g-$regular PDAs is provided which represents MADC models with NNC-MRGs in the same section. The GC-MRG is considered in Section \ref{GC-MRG} and it is proved that a set of $g-$regular MRAs corresponds to the existing scheme for MADC models with CT. Those  $g-$regular MRAs are extended to generate another set of MRAs to represent  MADC models with GC-MRGs in the same section along with providing a lower bound for MADC model with GC-MRGs under homogeneous setting.

{\it Notation:}  The bit wise exclusive OR (XOR) operation is denoted by $\oplus.$ The notation $[n]$ represents the set $\{1,2, \ldots , n\}$, $[a,b]$ represents the set $\{ a, a+1, \ldots, b \}$, while $[a,b)$ represents the set $\{a,a+1, \ldots , b-1\}$. $\lfloor x \rfloor$ denotes the largest integer smaller than or equal to $x$ and $\lceil x \rceil$ denotes the smallest integer greater than or equal to $x$. The notation
$a|b$ implies $a$ divides $b$, for some integers $a$ and $b$. For any $m \times n$ array  $\textbf{A} = (a_{i,j}),$ for $ i \in [0,m-1] $ and $n \in [0,n-1],$  the array $\textbf{A}+b$, is defined as $\textbf{A}+b =(a_{i,j}+b)$.
\section{Problem Definition}
\label{problem defintion}
\begin{figure*}
\begin{subfigure}{.5\textwidth}
	\centering
	\includegraphics[scale=0.55]{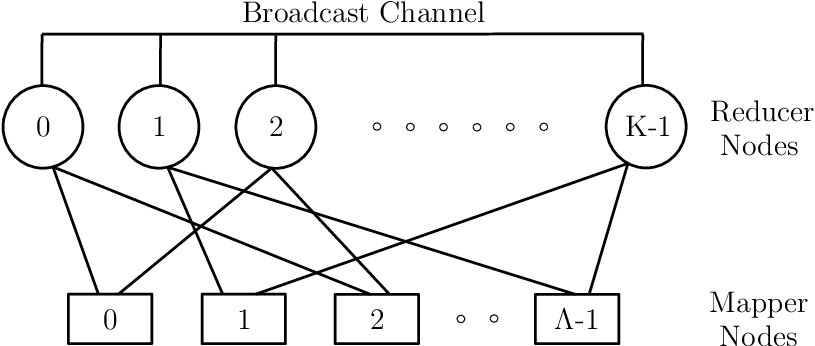}
	\caption{MADC Model.}
	\label{fig: madc model}
\end{subfigure}%
\begin{subfigure}{.5\textwidth}
	\centering
	\includegraphics[scale=0.53]{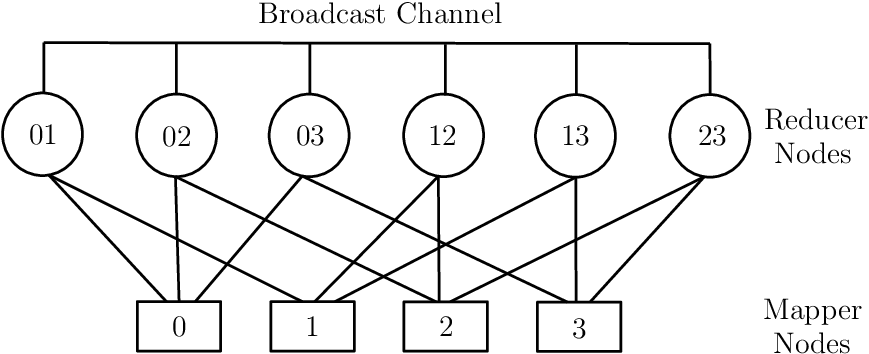}
	\caption{MADC model corresponding to Example \ref{exmp6}.}
	\label{fig: madc NNC-MRG exmp2}
\end{subfigure}
\end{figure*}
In the MADC model which has a Map-Reduce framework \cite{BP}, as shown as in Fig. \ref{fig: madc model}, there are $\Lambda$ mapper nodes indexed by $[0, \Lambda)$, and $K$ reducer nodes indexed by $[0,K)$. Each reducer node $k \in [0,K)$ is assigned to compute some output functions which depend on $N$ input files. Thus, the problem is to compute $Q$ output functions denoted as  $\{\phi_q:  q \in [0,Q)\}$, from $N$ input files denoted as $\{w_n : n \in [0,N)\}$, where the task is distributed across $K$ reducer nodes. Each file $w_n \in {\F}_{2^d}$ with $n \in [0,N)$ consists of $d$ bits and each function $\phi_q$ for $ q \in [0,Q)$ maps all  $N$ input files into a stream of $b$ bits, i.e., we have 
\begin{align}
\phi_q : {\F}_{2^d}^N \rightarrow {\F}_{2^b}.
\end{align}
 We assume that there is a map function $g_{q,n} : {\F}_{2^d} \rightarrow {\F}_{2^t}$ for each $n \in [0,N)$, which maps the input file $w_n$ into an intermediate value (IV) $v_{q,n} = g_{q,n}(w_n) \in {\F}_{2^t}$ of $t$ bits, and a reduce function, $h_q : {\F}_{2^t}^N \rightarrow {\F}_{2^b}$ which maps all IVs into the output value $h_q(v_{q,0}, \ldots , v_{q,N-1} ) \in {\F}_{2^b}$ of b bits. Thus, $\phi_q$ can be described as 
\begin{align}
\phi_q(w_0, \dots , w_{N-1}) = h_q(v_{q,0}, \ldots , v_{q,N-1} ),   \forall q \in [0,Q).
\end{align}

Each reducer node $k \in [0,K)$ is connected to  some mapper nodes and is assigned a subset of the output functions, $\mathcal{W}_k \subseteq [0,Q)$, where $\mathcal{W}_k$ contains the indices
of the functions assigned to the reducer node $k$. There is a symmetric assignment, which implies $|\mathcal{W}_k| = Q/K $ and $|\mathcal{W}_{k_1} \cap \mathcal{W}_{k_2}| = 0$, for all $k_1,k_2 \in [0,K)$ such that $k_1$ $ \neq k_2$.
The computation is carried out in three phases:
\begin{enumerate}
	\item {\bf Map Phase:}  The files are divided by grouping the $N$ files into $F$ disjoint batches $\mathcal{B} =\{B_{0},B_{1},\ldots, B_{F-1}\}$, each containing $\eta_1= N / F$ files such that $\bigcup_{m=0}^{F-1} B_{m} = \{w_0,w_1,\ldots,w_{N-1}\}$. Each mapper node $\lambda \in [0,\Lambda)$ locally stores a subset of batches 
	$M_{\lambda} \subseteq \mathcal{B}$
	and computes the set 
	\begin{align}
	\{v_{q,n} = g_{q,n}(w_n) : &q \in [0,Q), w_n \in B_{f},\nonumber\\&B_{f} \in M_{\lambda}, f \in [0,F)\}
	\end{align}
	 where each $v_{q,n}$ is a bit stream of length $t$ and is referred to as an IV.
	\item {\bf Shuffle Phase:} Each reducer node $k \in [0,K)$ is connected to some mapper nodes and can access all files which those mapper nodes have and retrieve the IVs from those mapper nodes. Each reducer node $k$ creates a sequence
	 \begin{align}
	 	 {\bf {X}}_k \in {\F}_{2^{l_k}}
	 \end{align}
	 and multicasts it to all other reducer nodes via the broadcast link which connects the reducer nodes. We assume that each reducer node receives all the multicast transmissions without any error.
	\item {\bf Reduce Phase:} Recall that each reducer $k \in [0,K)$ is assigned a subset of output functions whose indices are in $\mathcal{W}_k$ and requires to recover the IVs 
	\begin{align}
		\{v_{q,n} : q \in \mathcal{W}_k, n \in [0,N)\}
	\end{align}
	 to compute $\phi_q$, for each $q \in \mathcal{W}_k$. Receiving the sequences ${\{{\bf X}_j\}}_{j \in [0,K) \backslash k}$, each reducer node $k$ decodes all IVs $v_{q,n}$ of its output functions with the help of the IVs it has access to, and finally computes the output functions assigned to them.
\end{enumerate}
Like in CDC, the metrics which we consider for the evaluation of MADC models are computation and communication loads. Our objective is to optimize both of them.
\begin{defn}
	({\bf Computation Load} \cite{BP}): Computation load $r$ is defined as the total number of files mapped across the $\Lambda$ mapper nodes normalized by the total number of files, i.e., we have
	\begin{align}
	r := \frac{\sum_{\lambda=0}^{\Lambda-1} \eta_1|M_{\lambda}|}{N} =\frac{\sum_{\lambda=0}^{\Lambda-1} |M_{\lambda}|}{F}.
	\end{align}
\end{defn}
\begin{defn}
	({\bf Communication Load} \cite{BP}): The communication load $L$ is defined as the total number of bits transmitted by the $K$ reducer nodes over the broadcast channel during the Shuffle phase normalized by the number of bits of all IVs, i.e., we have
	\begin{align}
	L := \frac{\sum_{k \in [0,K)} l_k}{QNt}.
	\end{align}
\end{defn}
\subsection{BE Scheme \cite{BP}}
In \cite{BP}, the authors considered an MADC model with CT with $\Lambda$ mapper nodes and $K = {\Lambda \choose \alpha}$ reducer nodes, for fixed value $\alpha \in [\Lambda]$,  where there is exactly one reducer node for each
subset of $\alpha$ mapper nodes. 
A new scheme was proposed in \cite{BP}, which we refer as the \textit{BE scheme}. For a computation load of $r$, it is proved that the BE scheme allows for a coding gain of $g={r+\alpha \choose r}-1$, i.e., a coded transmission done by any reducer node benefits $g$ other reducer nodes during the shuffling phase, as compared to a maximal coding
gain of $r$ in the original CDC setting in \cite{LMA}. This is achieved by effectively utilizing the CT which in turn helps to reduce the communication load during the shuffling phase. The BE scheme achieves a communication load of $L_{BE}(r)$ which is a piecewise linear curve with corner points
	\begin{align}
	\label{BE rate}
	\left (r,L_{BE}(r)\right) = \left ( r, \frac{{\Lambda -\alpha\choose r}}{{\Lambda \choose r} \left ( {r+\alpha \choose r} -1\right )}\right), && \forall r \in [\Lambda-\alpha+1].
	\end{align}	
For this MADC model, a lower bound on the optimal communication load $L^{lb}_{BE}$ is derived in \cite{BP} which is a piecewise linear curve with corner points 
\begin{align}
	\label{BE lower bound}
	 \left (r,L^{lb}_{BE}(r)\right) = \left ( r, \frac{{\Lambda \choose r+\alpha}}{{\Lambda \choose r}{\Lambda \choose \alpha} }\right), &&&\forall r \in [\Lambda-\alpha+1].
\end{align}
\begin{example}
	\label{exmp6}
	Consider an MADC model with CT with $\Lambda = 4$ mapper nodes and $K = {\Lambda \choose \alpha}= 6$ reducer nodes,  i.e., $\{01, 02, 03, 12, 13,  23\}$, where $\alpha = 2$,
		as shown in Fig. \ref{fig: madc NNC-MRG exmp2}. 
	Assume that we have $N = 6$
	input files $\{w_0,w_1,w_2,w_3,w_4,w_5\}$ and $Q = 6$ output functions, $\{\phi_0,\phi_1,\phi_2,\phi_3,\phi_4,\phi_5\}$, to be computed across the reducer nodes. We assign $Q/K = 1$ output functions to each reducer node $U \in \{01, 02, 03, 12, 13, 23\}$. Let the indices of the output functions assigned to the reducer node $U$ be $\W_U$, where 
	\begin{align}
	\W_{\{01\}} &= \{0\},&
	\W_{\{02\}} &= \{1\},&
	\W_{\{03\}} &= \{2\},\nonumber\\
	\W_{\{12\}} &= \{3\},&
	\W_{\{13\}} &= \{4\},&
	\W_{\{23\}} &= \{5\}.
	\end{align}
	We partition $N = 6$ files into ${\Lambda \choose r} =6$ disjoint batches $B_{T}: T \in \{01,02,03,12,13,23\},$ where $r=2,$ i.e., we have
	\begin{align}
	B_{\{01\}} &= \{w_0\},&
	B_{\{02\}} &= \{w_1\},&
	B_{\{03\}} &= \{w_2\}, \nonumber\\
	B_{\{12\}} &= \{w_3\},&
	B_{\{13\}} &= \{w_4\},&
	B_{\{23\}} &= \{w_5\}.
	\end{align}
	For each $\lambda \in [0,4)$, mapper node $\lambda \in [0, 4)$ is assigned the set of files in $B_{T}$ if $\lambda \in T$, i.e., we have
	\begin{align}
	M_0 &= \{B_{\{01\}} , B_{\{02\}} , B_{\{03\}}\}, \nonumber\\
	M_1 &= \{B_{\{01\}} , B_{\{12\}} , B_{\{13\}} \}, \nonumber\\
	M_2 &= \{B_{\{02\}} , B_{\{12\}} , B_{\{23\}} \}, \nonumber\\
	M_3 &= \{B_{\{03\}} , B_{\{13\}} , B_{\{23\}}\} .
	\end{align}
	Each mapper node $\lambda$ computes $Q = 6$ intermediate values for each assigned input file. 
	The set of all files accessible to each reducer node $U$ is as follows:
	\begin{align}
	R_{\{01\}}  =\{B_{\{01\}},B_{\{02\}},B_{\{03\}},B_{\{12\}},B_{\{13\}}\} &= M_0 \cup M_1 \nonumber\\
	R_{\{02\}}  =\{B_{\{01\}},B_{\{02\}},B_{\{03\}},B_{\{12\}},B_{\{23\}}\} &= M_0 \cup M_2 \nonumber\\
	R_{\{03\}}  =\{B_{\{01\}},B_{\{02\}},B_{\{03\}},B_{\{13\}},B_{\{23\}}\} &= M_0  \cup M_3\nonumber\\
	R_{\{12\}}  =\{B_{\{01\}},B_{\{02\}},B_{\{12\}},B_{\{13\}},B_{\{23\}}\} &= M_1 \cup M_2\nonumber\\
	R_{\{13\}}  =\{B_{\{01\}},B_{\{03\}},B_{\{12\}},B_{\{13\}},B_{\{23\}}\} &= M_1 \cup M_3\nonumber\\
	R_{\{23\}}  =\{B_{\{02\}},B_{\{03\}},B_{\{12\}},B_{\{13\}},B_{\{23\}}\} &= M_2 \cup M_3
	\end{align}
	and each reducer node $U$ can retrieve all IVs in $V_U = \{v_{q,n} : q \in [0,6), w_n  \in B_{T}, B_{T}\in R_U\}$.
	
	We now desribe how each reducer node $U$ constructs its multicast message. Since the
	procedure is the same for all reducer nodes, we consider only	reducer node $\{0,1\}$ in this example.
		We let $\s = [0,4) \backslash \{0,1\} =\{2,3\}$. For each $\R \subseteq \s \cup \{0, 1\}$ such that
	$|\R| = 2$ and $\R \neq \{0,1\}$, and for $T_1 = (\s\cup\{0, 1\}) \backslash \R$, reducer node $\{0,1\}$ concatenates the IVs $\{v_{q,n} : q \in \W_{\R}, w_n\in B_{T_1} \}$ into the symbol $\U_{\W_{\R},B_{T_1}} = (v_{q,n} : q \in \W_{\R}, w_n\in B_{T_1})$. Notice
		that having $\R \neq \{0,1\}$ implies that $T_1 \cap \{0,1\} \neq \phi$, so reducer node $\{0,1\}$ can retrieve $B_{T_1}$ from the
		mapper nodes it is connected to and can construct the symbol $\U_{\W_{\R},B_{T_1}}$. Subsequently, such symbol is evenly split as
		\begin{align}
		\label{split}
		\U_{\W_{\R},B_{T_1}} = (\U_{\W_{\R},B_{T_1}}^{T_2}
		: T_2 \subseteq (\R \cup T_1), |T_2| = 2, T_2 \neq \R).
		\end{align}
This means that the reducer node $\{0, 1\}$ creates the symbols
\begin{align}
\U_{\W_{\{02\}},B_{\{13\}}} &= (v_{q,n} : q \in \W_{\{02\}}, w_n \in B_{\{13\}}) \nonumber\\
\U_{\W_{\{03\}},B_{\{12\}}} &= (v_{q,n} : q \in \W_{\{03\}}, w_n \in B_{\{12\}}) \nonumber\\
\U_{\W_{\{12\}},B_{\{03\}}} &= (v_{q,n} : q \in \W_{\{12\}}, w_n \in B_{\{03\}}) \nonumber\\
\U_{\W_{\{13\}},B_{\{02\}}} &= (v_{q,n} : q \in \W_{\{13\}}, w_n \in B_{\{02\}})  \nonumber\\
\U_{\W_{\{23\}},B_{\{01\}}} &= (v_{q,n} : q \in \W_{\{23\}}, w_n \in B_{\{01\}}) 
\end{align}
	
Each of the symbols above is then split into $5$ segments as (\ref{split}).
	Each reducer node $U$ constructs one coded message as
	\begin{align}
	X_0^U = \bigoplus\limits_{\R \subseteq (\s \cup U) : |\R| =\alpha, \R \neq U} \U_{\W_{\R}, (\s \cup U) \backslash \R}^U
	\end{align}
	 The following are the coded symbols transmitted by the reducer nodes.
	
	{\small
		\begin{align}
		X_0^{\{01\}} = &\U_{\W_{\{23\}},B_{\{01\}}}^{\{01\}} \oplus \U_{\W_{\{13\}},B_{\{02\}}}^{\{01\}} \oplus \U_{\W_{\{12\}},B_{\{03\}}}^{\{01\}} \nonumber \\&\oplus \U_{\W_{\{03\}},B_{\{12\}}}^{\{01\}} \oplus \U_{\W_{\{02\}},B_{\{13\}}}^{\{01\}} ,\nonumber\\
		X_0^{\{02\}} = &\U_{\W_{\{23\}},B_{\{01\}}}^{\{02\}} \oplus \U_{\W_{\{13\}},B_{\{02\}}}^{\{02\}} \oplus \U_{\W_{\{12\}},B_{\{03\}}}^{\{02\}} \nonumber \\&\oplus \U_{\W_{\{03\}},B_{\{12\}}}^{\{02\}} \oplus \U_{\W_{\{01\}},B_{\{23\}}}^{\{02\}} ,\nonumber\\
		X_0^{\{03\}} = &\U_{\W_{\{23\}},B_{\{01\}}}^{\{03\}} \oplus \U_{\W_{\{13\}},B_{\{02\}}}^{\{03\}} \oplus \U_{\W_{\{12\}},B_{\{03\}}}^{\{03\}} \nonumber \\&\oplus \U_{\W_{\{02\}},B_{\{13\}}}^{\{03\}} \oplus \U_{\W_{\{01\}},B_{\{23\}}}^{\{03\}} ,\nonumber\\
		X_0^{\{12\}} = &\U_{\W_{\{23\}},B_{\{01\}}}^{\{12\}} \oplus \U_{\W_{\{13\}},B_{\{02\}}}^{\{12\}} \oplus \U_{\W_{\{03\}},B_{\{12\}}}^{\{12\}} \nonumber \\&\oplus \U_{\W_{\{02\}},B_{\{13\}}}^{\{12\}} \oplus \U_{\W_{\{01\}},B_{\{23\}}}^{\{12\}} ,\nonumber\\
		X_0^{\{13\}} = &\U_{\W_{\{23\}},B_{\{01\}}}^{\{13\}} \oplus \U_{\W_{\{12\}},B_{\{03\}}}^{\{13\}} \oplus \U_{\W_{\{03\}},B_{\{12\}}}^{\{13\}} \nonumber \\&\oplus \U_{\W_{\{02\}},B_{\{13\}}}^{\{13\}} \oplus \U_{\W_{\{01\}},B_{\{23\}}}^{\{13\}} ,\nonumber\\
		X_0^{\{23\}} = &\U_{\W_{\{13\}},B_{\{02\}}}^{\{23\}} \oplus \U_{\W_{\{12\}},B_{\{03\}}}^{\{23\}} \oplus \U_{\W_{\{03\}},B_{\{12\}}}^{\{23\}} \nonumber \\&\oplus \U_{\W_{\{02\}},B_{\{13\}}}^{\{23\}} \oplus \U_{\W_{\{01\}},B_{\{23\}}}^{\{23\}} .
		\label{trans}
		\end{align}}
	The reducer node $\{01\}$ can retrieve $\U_{\W_{\{01\}},B_{\{23\}}}^{\{02\}}$  from the coded symbol $X_0^{\{02\}}$ transmitted by the reducer node $\{02\}$, since it can compute the rest of the symbols from the files in $B_{\{01\}}$.
	Similarly reducer node $\{01\}$ can retrieve $\U_{\W_{\{01\}},B_{\{23\}}}^{\{03\}}, \U_{\W_{\{01\}},B_{\{23\}}}^{\{12\}},\U_{\W_{\{01\}},B_{\{23\}}}^{\{13\}},$ and $\U_{\W_{\{01\}},B_{\{23\}}}^{\{23\}}$ from $X_0^{\{03\}}, X_0^{\{12\}},X_0^{\{13\}},$ and $X_0^{\{23\}}$ respectively. Hence, reducer node $\{01\}$ can compute the function $\phi_0$.
	It can be verified that all other reducer nodes can retrieve all required symbols needed to compute the respective functions.
	
	A total of $6$ coded symbols are transmitted, each with length $\frac{t}{5}$ bits. Hence the communication load is $L_{BE}(2) =\frac{t*6}{5*6*6*t} =\frac{1}{30}$.
\end{example}
\subsection{Placement Delivery Array \cite{YCTCPDA}}
Placement Delivery Array (PDA) was introduced by Yan et al. \cite{YCTCPDA} to represent the coded caching schemes with an aim to reduce sub-packetization level. The concept of PDA has been identified as an effective tool to reduce the sub-packetization level and since then various coded caching schemes based on the concept of PDA were reported.
\begin{defn}  ({\bf Placement Delivery Array}\cite{YCTCPDA}):
	For positive integers $K, F, Z,$ and $S,$ an $F \times K$ array $P = [p_{f,k}]$ with $ f \in [0,F),$ and $ k \in [0,K)$ composed of a specific symbol $*$ and $S$ non-negative integers $[0,S),$ is called a $(K, F, Z, S)$ Placement Delivery Array (PDA) if it satisfies the following conditions:
	\begin{itemize}
		\item {\it A1:} The symbol $*$ appears $Z$ times in each column;
		\item {\it A2:} Each integer occurs at least once in the array;
		\item {\it A3:} For any two distinct entries $p_{f_1,k_1}$ and $p_{f_2,k_2}, s=p_{f_1,k_1} = p_{f_2,k_2} $ is an integer only if
		\begin{enumerate}
			\item $f_1$ $\neq f_2$ and $ k_1$ $\neq k_2,$ i.e., they lie in distinct rows and distinct columns; and
			\item $p_{f_1,k_2} = p_{f_2,k_1} = *,$ i.e., the corresponding $2 \times 2$ sub-array formed by rows $f_1, f_2$ and columns $k_1, k_2$ must be either of the following forms
			$ \begin{pmatrix}
			s & *\\
			* & s
			\end{pmatrix} $or 
			$\begin{pmatrix}
			*& s\\
			s & *
			\end{pmatrix}.$
		\end{enumerate}
	\end{itemize}
\label{def:PDA}
\end{defn}
\begin{example}
	\label{PDA example}
	Consider an $4 \times 4$ array $A_1$ as given below.
	\begin{equation}
	A_1 =
	\begin{blockarray}{cccc}
	\begin{block}{(cccc)}
	* & * & * & 0    \\
	* & 0 & 1 & *   \\
	0 & * & 2 & *   \\
	1 & 2 & * & 3   \\
	\end{block}
	\end{blockarray}
	\end{equation}
	The array $A_1$ satisfies conditions \textit{A1, A2} and \textit{A3}. There are $2$ stars in each column and a total of $4$ integers in the array. Hence, the array $A_1$ is a $(4,4,2,4)$ PDA.
\end{example}
\begin{defn} ({\bf $g-$regular PDA}\cite{YCTCPDA}):
	An array $P$ is said to be a $g-(K, F, Z, S)$ PDA if it satisfies C1, C3, and the following condition
	\begin{itemize}
		\item {\it $A2'$:}  Each integer appears $g$ times in $P$, where $g$ is a constant.
	\end{itemize}
\label{def:g-PDA}
\end{defn}
\begin{example}
	\label{g regular example}
	The $4 \times 6$ array $A_2$ provided below is a $3-(6,4,2,4)$ PDA.
	\begin{equation}
	A_2 =
	\begin{blockarray}{cccccc}
	\begin{block}{(cccccc)}
	* & * & * & 0 &  1&2  \\
	* & 0 & 1 & * & *& 3 \\
	0 & * & 2 & *  &  3 &*\\
	1 & 2 & * & 3 & *  &*\\
	\end{block}
	\end{blockarray}
	\end{equation}
\end{example}
\begin{defn}\cite{SR} ({\bf$l$-cyclic $g$-regular  PDA}): \label{def new PDA}
	In a $g-(K,F,Z,S)$ PDA P, if all the $Z$ stars in each column occur in consecutive rows and if the position of stars in each column in P is obtained by cyclically shifting the previous column downwards by $l$ units, then it is called as \textit{$(l,g)-(K,F,Z,S)$ PDA}. 
	\label{def:l-g-PDA}
\end{defn}
\begin{example}
	\label{l g regular example}
	The $4 \times 4$ array $A_3$ provided below is a $(2,2)-(4,4,2,4)$ PDA.
	\begin{equation}
	A_3 =
	\begin{blockarray}{cccccc}
	\begin{block}{(cccccc)}
	* & 0 & * & 2   \\
	* & 1 & * & 3  \\
	0 & * & 2 & *  \\
	1 & * & 3 & * \\
	\end{block}
	\end{blockarray}
	\end{equation}
\end{example}
\section{Map-Reduce Graphs and Map-Reduce Arrays}
\label{MRA}	
\begin{figure}[!t]
		\centering
		\includegraphics[scale=0.6]{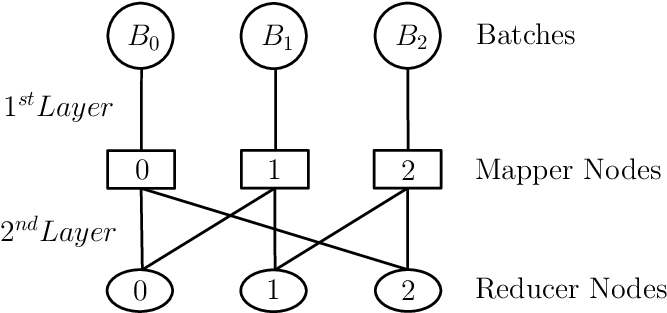}
		\caption{MRG for MADC model consisting of $3$ batches of files, mapper nodes and reducer nodes with $r=1$ and each reducer node connected to $2$ mapper nodes.}
		\label{fig: eg1}
	\end{figure}

In this section,  we define a 2-layered bipartite graph named as Map-Reduce Graph to represent MADC models, where the association of batches, mapper nodes and reducer nodes can be visualized. Moreover, we define a new array inspired from PDA by relaxing one of the conditions of the PDA, namely condition \textit{A1} in Definition \ref{def:PDA}, and modifying \textit{A2}.  We call it as Map-Reduce Array.
\begin{defn}
	(\textbf{Map-Reduce Graph}): An MADC model is visualized using a 2-layered bipartite graph featuring three sets of vertices. The first set represents  batches of files, the second represents mapper nodes, and the last represents reducer nodes. In the first layer, the graph illustrates the association of mapper nodes with batches, i.e., a vertex representing a mapper node is connected to a vertex representing a batch if and only if the mapper node has access to that batch. The second layer reveals the connection between reducer nodes and mapper nodes, i.e., there is an edge between a vertex representing a reducer node and a vertex representing a mapper node if and only if the reducer node is connected to that mapper node. This graphical representation is termed as a Map-Reduce Graph (MRG).
\end{defn} 
 It can be observed that the computation load $r$ is the sum of the degrees of the vertices representing the batches divided by the total number of vertices representing the batches. Thus the value $r$ can be derived from MRGs.
\begin{example}
	Consider the MRG provided in Fig. \ref{fig: eg1}. There are three sets of vertices representing batches, mapper nodes and reducer nodes. Each set contains $3$ vertices, i.e., there are $3$ batches of files, mapper nodes and reducer nodes in the corresponding MADC model. Each mapper node $\lambda \in \{0,1,2\}$ has access to the batch $B_{\lambda}$ and each reducer node $k \in \{0,1,2\}$ is connected to the mapper nodes $k$ and $(k+1)$ mod $3$. This is the MADC model corresponding to the MRG in  Fig. \ref{fig: eg1}.
\end{example}
\begin{defn} ({\bf Map-Reduce Array}):
	For positive integers $K, F,$ and $S,$ an $F \times K$ array $P = [p_{f,k}]$ with $ f \in [0,F),$ and $ k \in [0,K)$ composed of a specific symbol $*$ and $S$ non-negative integers $[0,S),$ is called a $(K, F, S)$ Map-Reduce Array (MRA) if it satisfies the following conditions: 
	\label{def:GPDA}
	\begin{itemize}
		\item {\it C1:} Each integer occurs more than once in the array;
		\item {\it C2:} For any two distinct entries $p_{f_1,k_1}$ and $p_{f_2,k_2}, s=p_{f_1,k_1} = p_{f_2,k_2} $ is an integer only if
		\begin{enumerate}
			\item $f_1$ $\neq f_2$ and $ k_1$ $\neq k_2,$ i.e., they lie in distinct rows and distinct columns; and
			\item $p_{f_1,k_2} = p_{f_2,k_1} = *,$ i.e., the corresponding $2 \times 2$ sub-array formed by rows $f_1, f_2$ and columns $k_1, k_2$ must be either of the following forms
			$ \begin{pmatrix}
			s & *\\
			* & s
			\end{pmatrix} $or 
			$\begin{pmatrix}
			*& s\\
			s & *
			\end{pmatrix}.$
		\end{enumerate}
	\end{itemize}
\end{defn}
\begin{defn}  ({\bf $g-$regular MRA}):
	An array $P$ is said to be a $g-$regular
	$(K, F, S)$ MRA if it satisfies {\it C2}, and the following condition
	\begin{itemize}
		\item {\it $C1'$:}  Each integer appears $g$ times in $P$, where $g \geq 2$ is a constant.
	\end{itemize}
	\label{def:g-MRA}
\end{defn}
\begin{example}
	\label{MRA example}
	Consider an $4 \times 5$ array $P_1$ as given below.
	\begin{equation}
	P_1 =
	\begin{blockarray}{ccccc}
		\begin{block}{(ccccc)}
			 * & * & * & 0 & *   \\
			* & 0 & 1 & * & 3  \\
			 0 & * & 2 & *  & *  \\
			 1 & 2 & * & 3 & *  \\
		\end{block}
	\end{blockarray}
\end{equation}
The array $P_1$ satisfies conditions \textit{C1} and \textit{C2}. There are $4$ integers in the array. Hence, the array $P_1$ is a $(5,4,4)$ MRA.
\end{example}
\begin{example}
	\label{g regular MRA example}
	An example for $2-$regular $(4,5,5)$ MRA is given below.
	\begin{equation}
	P_2=	\begin{blockarray}{cccc}
			\begin{block}{(cccc)}
				* & 0 & 2 & *   \\
				* & 1 & * & 3   \\
				0 & * & 3 & *   \\
				1 & * & 4 & *   \\
				2 & * & * & 4   \\
			\end{block}
		\end{blockarray}
	\end{equation}
\end{example}
\begin{remark}
	\label{remark1}
	It can be observed that given an $(K, F, Z, S)$ PDA P, for positive integers $K, F, Z,$ and $S,$ if each integer occurs more than once in the PDA, then the PDA P is a $(K,F,S)$ MRA. The reverse is not always true. An $(K,F,S)$ MRA is a  $(K, F, Z, S)$ PDA P only if it obeys {\it A1}. Also, all $g-$regular PDAs with $g\geq 2$ are $g-$regular MRAs but all $g-$regular MRAs are not $g-$regular PDAs. The MRAs illustrated in Examples \ref{MRA example} and \ref{g regular MRA example} are not PDAs.
\end{remark}
\noindent We state one of the important properties of MRA below.
\begin{property}
	An array $\hat{P}$ obtained by removing some of the columns in an 
	$(K, F, S)$ MRA for positive integers $K, F,$ and $S,$ is also an MRA as long as condition {\it C1} holds, i.e., as long as each integer in  $\hat{P}$ appears more than once.
\end{property}
\begin{figure}[!t]
	\centering
	\includegraphics[scale=0.6]{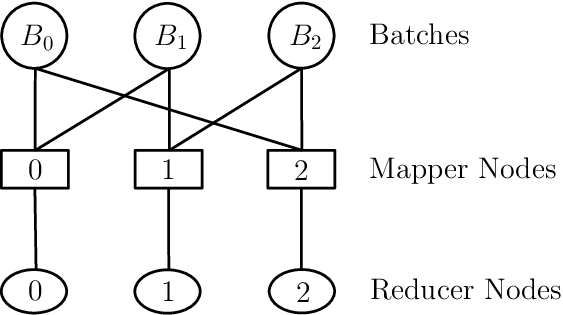}
	\caption{MRG for MADC model consisting of $3$ batches of files, mapper nodes and reducer nodes with $r=2$ and each reducer node connected to $1$ mapper nodes.}
	\label{fig: eg2}
\end{figure}
\begin{figure}[!t]
	\centering
	\includegraphics[scale=0.6]{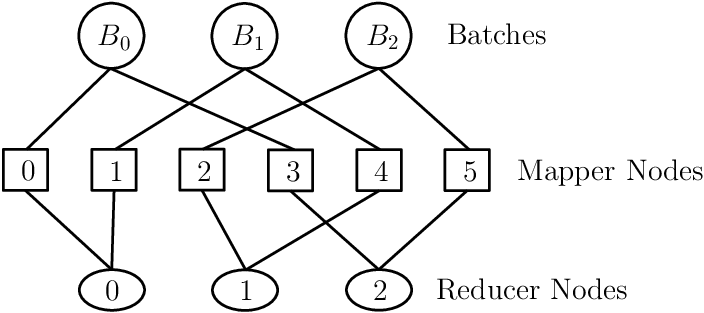}
	\caption{MRG for MADC model consisting of $3$ batches of files, $6$ mapper nodes and $3$ reducer nodes with $r=2$ and each reducer node connected to $2$ mapper nodes.}
	\label{fig: eg3}
\end{figure}
\begin{example}
	\label{MRA property eg}
	Consider an $4 \times 3$ array $\hat{P}_1$ obtained by removing the last two columns in $P_1$ as shown below.
	\begin{equation}
		\hat{P}_1 =
		\begin{blockarray}{ccc}
			\begin{block}{(ccc)}
				* & * & *    \\
				* & 0 & 1 \\
				0 & * & 2  \\
				1 & 2 & *   \\
			\end{block}
		\end{blockarray}
	\end{equation}
	The array $	\hat{P}$ satisfies conditions \textit{C1} and \textit{C2}. Hence, it is an $(3,4,3)$ MRA. If we remove just the last column from $P_1$, then the truncated array is not an MRA as the integer $3$ is only appearing once in the truncated array.
\end{example}
\section{Topologies from MRAs}
\label{NT}	
 In this section, we connect MRAs to MADC models. In Theorem \ref{thm1}, for a given MRA, if an MADC model obeys certain conditions, then, we provide a coding scheme based on the structure of MRAs. The proof of Theorem \ref{thm1} is provided in Appendix \ref{proof thm1}.
\begin{thm}
	\label{thm1}
	Consider that we are given an $(K, F,S)$ MRA $P= [p_{f,k}]$ for $ f\in [0,F),k\in[0,K)$, and for some integers $ K,F$ and $S$. Suppose an  MADC model has $K$ reducer nodes, indexed by $[0,K)$, and $F$ batches of files, indexed by $\{B_0,B_1,\ldots,B_{F-1}\}$. For some $\Lambda$ mapper nodes and computation load $r$, if each mapper node is assigned a subset of batches and connected to some reducer nodes in such a way that each reducer node $k \in [0,K)$ can access all batches in the set
		\begin{align}
			\label{reducer config}
		R_k = \{B_{f} : p_{f,k} = *, f \in [0,F)\}
		\end{align}
	then, a communication load achievable for the corresponding MADC model is given by
	\begin{align}
	\label{comm}
	L(r)= \frac{S}{KF}+\sum_{g=2}^{K}\frac{S_g}{KF (g-1)}
	\end{align}	
	\noindent where $S_g$ is the number of integers in $[0,S)$ which appears exactly $g$ times in the MRA $P$. 
\end{thm}
It is observed from Theorem \ref{thm1} that given an $(K, F,S)$ MRA, the row index $f \in [0,F)$ represents the batch $B_{f}$ and the column index $k \in [0,K)$ represents the reducer node $k$. There exists a $*$ in a row indexed by $B_f$ and column indexed by $k$ if and only if the reducer node $k$ has access to the batch $B_f$, for each $f \in [0,F)$ and $k \in [0,K)$. We can have multiple MADC models that correspond to a given MRA since we do not put restriction on $r$ or $\Lambda$. This is illustrated in the following example.
\begin{example}
	\label{eg comparison}
	Consider a $3-$regular $(3,3,1)$ MRA given below.
	\begin{equation}
	\label{simple mra example}
	\begin{blockarray}{cccc}
	& \{0\} & \{1\} & \{2\}  \\
	\begin{block}{c(ccc)}
	B_{0} & * & 0 & *   \\
	B_{1} &* & * & 0  \\
	B_{2} & 0 & * & * \\
	\end{block}
	\end{blockarray}
	\end{equation}
	The row index $f \in [0,3)$ represents the batch $B_{f}$ and the column index $k \in [0,3)$ represents the reducer node $k$. There exists a $*$ in a row indexed by $B_f$ and column indexed by $k$ if and only if the reducer node $k$ has access to the batch $B_f$, for each $f \in [0,3)$ and $k \in [0,3)$.
	The MRGs provided in Figures \ref{fig: eg1}, \ref{fig: eg2}, and \ref{fig: eg3} correspond to this MRA. The number of batches and reducer nodes in all the MRGs are same. The MRGs in Figures \ref{fig: eg1} and \ref{fig: eg2} have the same number of mapper nodes (the value of $\Lambda$) while the computation load and the number of mapper nodes to which each reducer node is connected to differ. The MRGs in Figures \ref{fig: eg2} and \ref{fig: eg3} have the same computation load while the value of $\Lambda$ and the number of mapper nodes to which each reducer node is connected to is different. For all the MRGs, by Theorem \ref{thm1}, the communication load is same. Since our aim is to minimize the computation load as well, we prefer the MRG with the least computation load, i.e., $r=1$ (MRG in Fig. \ref{fig: eg1}).
\end{example}
	Based on the insights from Example \ref{eg comparison}, given an MRA, we derive the MADC model with the least computation load, i.e., $r=1$ in Theorem \ref{thm2}.
	The proof of Theorem \ref{thm2} is provided in Appendix \ref{proof thm2}.
\begin{thm}
	\label{thm2}
	 Given an $(K, F,S)$ MRA $P= [p_{f,k}]$ for $ f\in [0,F),k\in[0,K)$, and for some integers $ K,F$ and $S$, there exists a coding scheme for an MADC model with the MRG which consists of 
	\begin{enumerate}
		\item $F$ batches of files, indexed by $\{B_0,B_1,\ldots,B_{F-1}\}$;
		\item $F$ mapper nodes, indexed by $[0,F)$;
	\item $K$ reducer nodes, indexed by $[0,K)$;
	\item Each mapper node $f \in [0,F)$  has access to the batch $B_f$, and;
		\item Each reducer node $k \in [0,K)$  is connected to mapper nodes in the set
		$ \{f : p_{f,k} = *, f \in [0,F)\}.$
	\end{enumerate}
	For the corresponding MADC model, the computation load is $r=1$ and the communication load achievable is given by
		\begin{align}
		L(1)= \frac{S}{KF}+\sum_{g=2}^{K}\frac{S_g}{KF (g-1)}
		\end{align}	
	\noindent where $S_g$ is the number of integers in $[0,S)$ which appears exactly $g$ times in the MRA $P$. 
\end{thm}
\begin{remark}
	It is observed from Theorem \ref{thm2} that the number of mapper nodes to which each reducer node $k \in [0,K)$ is connected to is given by the number of the symbol $*$ in the column indexed by $k$ of the MRA. In an MRA, the number of $*$ in each column need not be the same. Hence, the number of mapper nodes to which each reducer node is connected to can vary depending upon the MRA. If the $(K,F,S)$ MRA is a $(K,F,Z,S)$ PDA, for some positive integers $K,F,Z$ and $S$, then, we know that the number of $*$ in each column is $Z$ and hence, in the corresponding MADC model, each reducer node is connected to some $Z$ mapper nodes.
\end{remark}
\noindent We consider an example to illustrate  Theorem \ref{thm2}.
\begin{figure}[!t]
	\centering
	\includegraphics[scale=0.56]{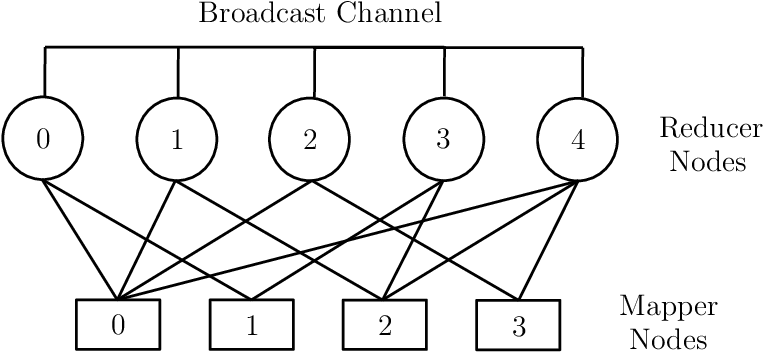}
	\caption{MADC model corresponding to Example \ref{exmp1} with $4$ mapper nodes, $5$ reducer nodes.}
	\label{fig: madc NNC-MRG exmp1}
\end{figure}
\begin{example}
	\label{exmp1}
	Consider an MADC model where there are $N = 4$ input files $\{w_0,w_1,w_2,w_3\}$, and $Q = 5$ output functions $\{\phi_0, \phi_1,\phi_2,\phi_3,\phi_4\}$ to be computed. Consider the $(5, 4,  4)$ MRA $P_1$ of Example \ref{MRA example}.
	\begin{equation}
	\label{general pda example}
	P_1 =
	\begin{blockarray}{cccccc}
	& \{0\} & \{1\} & \{2\} &  \{3\}& \{4\} \\
	\begin{block}{c(ccccc)}
	B_{0} & * & * & * & 0 & *   \\
	B_{1} &* & 0 & 1 & * & 3  \\
	B_{2} & 0 & * & 2 & *  & *  \\
	B_{3} & 1 & 2 & * & 3 & *  \\
	\end{block}
	\end{blockarray}
	\end{equation}
 The row index $f \in [0,4)$ represents the batch $B_{f}$ and the column index $k \in [0,5)$ represents the reducer node $k$. There exists a $*$ in a row indexed by $B_f$ and column indexed by $k$ if and only if the reducer node $k$ has access to the batch $B_f$, for each $f \in [0,4)$ and $k \in [0,5)$.
	Consider that there are $\Lambda = F=4$ mapper nodes indexed by $\{0,1,2,3\}$, $K = 5$ reducer nodes indexed by $\{0,1,2,3,4\}$, and we partition $N = 4$ files into $4$ batches $\{B_{0},B_{1},B_{2},B_{3}\}$, where $B_{f} =\{w_{f}\} $ for $ f \in [0,4)$. Let us assign $Q/K = 1$ output function to each reducer node $k \in \{0,1,2,3,4\}$. Let the indices of the output functions assigned to the reducer node $k$ be $\mathcal{W}_{k}= \{k\}.$
	The batch assigned to mapper node $f \in [0, 4)$ is given by  $M_{f} = \{B_f\}$. Hence, the computation load is $r=1.$
	For each $f \in [0,4)$, the mapper node $f$ computes $Q = 5$ IVs for each assigned input file. 
	
	Consider an MADC model where each reducer node $k$ is connected to the mapper nodes in the set $\{f: p_{f,k} = *, f  \in [0,4)\}$ as shown in Fig \ref{fig: madc NNC-MRG exmp1}. It can be observed that the array $P_1$ corresponds to this model since the set of all batches assigned to each reducer node $k$ is as follows:
	\begin{align}
	R_{k}  =  \bigcup_{f\in [0,4):p_{f,k} = *} M_{f}.
	\end{align}
	Each reducer node $k$ can retrieve all the IVs in $V_k = \{v_{q,n} : q \in [0,5), w_n \in B_{f} , B_{f}\in R_{k}, f \in [0,4)\}$.  	
	Consider the first column, i.e. column with index $0$ of $P_1$. The set of all integers present in this column is $S_{0} = \{0,1\}$. We concatenate the IVs for the output functions in $\mathcal{W}_{0}$ which need to be computed by the reducer node $0$ and can be computed from the files in $B_{2}$, i.e., $\{v_{q,n} : q \in \W_{0}, w_n \in B_{2} \}$, into a symbol
	\begin{align}
	\U_{\W_{0},B_{2}} = (v_{q,n} : q \in \{0\}, w_n \in \{w_2\}).
	\end{align}
	Similarly, we concatenate the IVs for the output functions in $W_{0}$ which need to be computed by the reducer node $0$ and can be computed from the files in $B_{3}$, i.e., $\{v_{q,n} : q \in \W_{0}, w_n \in B_{3} \}$, into the symbol 
	\begin{align}
	\U_{\W_{0},B_{3}} = (v_{q,n} : q \in \{0\}, w_n \in \{w_3\}).
	\end{align}
	Consider the entry $s= 0 $ in $ S_{0}$. The other entries which are $0
	$ are in the columns indexed by $1$ and $3$. Hence, we partition the symbols in $\U_{\W_{0},B_{2}}$ into $2$ packets, each of equal size, 
	\begin{align}
	\U_{\W_{0},B_{2}} =\{\U_{\W_{0},B_{2}}^{1}, \U_{\W_{0},B_{2}}^{3} \} .
	\end{align}
	Next for the entry $s= 1$ in $ S_{0}$, we partition $\U_{\W_{0},B_{3}}$ into $1$ packet which is the symbol itself, since the other entry which is $1$ correspond to  the column $2$.
	\begin{align}
	\U_{\W_{0},B_3} =\{\U_{\W_{0},B_{3}}^{2}\}.
	\end{align}
	Similarly, for each column $k \in [1,5)$, we concatenate the IVs for the output functions in $\W_{k}$, which need to be computed by the reducer node $k$ and can be computed from the files not accessible to them, as
	\begin{align}
	\U_{\W_{k},B_{j}} &= (v_{q,n} : q \in \{k\}, w_n \in \{w_{j}\}),
	\end{align}
	where $j \in [0,4)$ such that $p_{j,k} \neq *$.
	For entry  $0$ we partition the corresponding symbol into $2$ packets of equal sizes, while for entries $1,2$ and $3$, we partition the symbols into $1$ packet which is the symbol itself. The partitioning is shown below.
	\begin{align}
	\U_{\W_{1},B_{1}} &= \{\U_{\W_{1},B_{1}}^{0},\U_{\W_{1},B_{1}}^{3}\},&
	\U_{\W_{1},B_{3}} &= \{\U_{\W_{1},B_{3}}^{2}\}, \nonumber\\
	\U_{\W_{2},B_{1}} &= \{\U_{\W_{2},B_{1}}^{0}\},&
	\U_{\W_{2},B_{2}} &= \{\U_{\W_{2},B_{2}}^{1}\}, \nonumber\\
	\U_{\W_{3},B_{0}} &= \{\U_{\W_{3},B_{0}}^{0},\U_{\W_{3},B_{0}}^{1}\},&
	\U_{\W_{3},B_{3}} &= \{\U_{\W_{3},B_{3}}^{4}\}, \nonumber\\
	\U_{\W_{4},B_{0}} &= \{\U_{\W_{4},B_{0}}^{0},\U_{\W_{4},B_{0}}^{2}\},&
	\U_{\W_{4},B_{1}} &= \{\U_{\W_{4},B_{1}}^{3}\}.
	\end{align}
	Let $S_k$ denote the set of all integers present in column $k$, for  $k\in [0,5)$. Since $|S_k|=2, \forall k\in [0,4)$, each reducer node $k$ transmits two coded symbols $X_s^k$ for $s\in S_{k}$, while reducer node $4$ transmits one symbol $X_3^4$. The following are the coded symbols transmitted by the reducer nodes.
	\begin{align}
	X_0^{0} &= \U_{\W_{1},B_{1}}^{0}  \oplus \U_{\W_{3},B_{0}}^{0}, &
	X_1^{0} &= \U_{\W_{2},B_{1}}^{0} , \nonumber\\
	X_0^{1} &= \U_{\W_{0},B_{2}}^{1}  \oplus \U_{\W_{3},B_{0}}^{1}, &
	X_2^{1} &= \U_{\W_{2},B_{2}}^{1} , \nonumber\\
	X_1^{2} &= \U_{\W_{0},B_{3}}^{2} , &
	X_2^{2} &= \U_{\W_{1},B_{3}}^{2}, \nonumber\\
	X_0^{3} &= \U_{\W_{0},B_{2}}^{3}  \oplus \U_{\W_{1},B_{1}}^{3}, &
	X_3^{3} &= \U_{\W_{4},B_{1}}^{3} , \nonumber\\
	X_3^{4} &= \U_{\W_{3},B_{3}}^{4} .
	\end{align}
	The reducer node $0$ can retrieve $\U_{\W_{0},B_{2}}^{1}$  from the coded symbol $X_0^{1}$ transmitted by the reducer node $1$, since it can compute $\U_{\W_{3},B_{0}}^{1}$ from the files in $B_{0}$. Similarly, it can retrieve $\U_{\W_{0},B_{2}}^{3}$ and  $\U_{\W_{0},B_{3}}^{2}$ as well from the coded symbol $X_0^{3}$ and $X_1^{2}$ respectively.
	Hence, the reducer node $0$ can compute the function $\phi_0$. It can be verified that all other nodes can retrieve all required symbols needed to compute the respective functions. A total of $9$ coded symbols are transmitted across the reducer nodes. The symbols corresponding to the entry $0$ in the array (\ref{general pda example}) are of size $\frac{t}{2}$ bits while  symbols corresponding to the entries $1,2$ or $3$ are of size $t$ bits.  Hence, the communication load is $L(1) =\frac{\frac{t}{2} *3+t*6}{5*4*t} =\frac{15}{40}$.
\end{example}
\section{Nearest Neighbor Connect-MRG}
\label{NNC-MRG}
From Remark \ref{remark1}, we observe the connection between PDAs and MRAs. In this section, we explore some existing PDAs which satisfies condition {\it C1} and connect those to MADC models with certain MRGs.
\begin{defn}
	({\bf Nearest Neighbor Connect MRG}): Consider an MADC model with $\Lambda$ batches of files, mapper nodes and reducer nodes. An MRG is said to be Nearest Neighbor Connect-MRG (NNC-MRG), if it satisfies the following two conditions:
	\begin{enumerate}
		\item Each mapper node has access to $r$ consecutive batches in a sequential way, i.e., each mapper node $\lambda \in [0,\Lambda)$ has access to the following batches of files:
		\begin{align}
		M_{\lambda} = \{B_{(r\lambda+j) \text{ mod }\Lambda}: j \in [0,r)\},
		\end{align}
		for some $r \in [\Lambda]$.
		\item Each reducer node $\lambda \in [0,\Lambda)$ is connected to $\alpha$  neighboring mapper nodes in a cyclic wrap-around way, i.e. reducer node $\lambda$ is connected to the mapper nodes in the set $\{\lambda+j: j\in [0,\alpha)\}$, for some $\alpha \in [\Lambda]$.
	\end{enumerate}
\end{defn}
For NNC-MRGs, in the map phase, we split the $N$ files into $\Lambda$ batches, $\{B_0,B_1,\ldots ,B_{\Lambda-1}\}$. Each mapper node $\lambda \in [0,\Lambda)$ is filled with batches of files as follows:
\begin{align}
M_{\lambda} = \{B_{(r\lambda+j) \text{ mod }\Lambda}: j \in [0,r)\}
\end{align}
Each mapper node stores $r$ batches of files. Hence the computation load is $\frac{r\Lambda}{\Lambda} =r$. 

The mapping is done in such a way that we first create a
list of size $1\times r\Lambda$ by repeating the sequence $\{B_0, B_1, \ldots, B_{\Lambda -1}\}$, $r$ times, i.e., $\{B_0,B_1,\ldots,B_{\Lambda-1},B_0,B_1,\dots,B_{\Lambda-1}, \ldots\}$. We map the  batches by sequentially taking from the list. Hence, the first mapper node is mapped with the first $r$ items, the second mapper node with the next $r$ items and so on.

Each reducer node can access $\alpha$ mapper nodes and each mapper node has  $r$ consecutive batches of files. If $\alpha \geq \left \lceil \frac{\Lambda}{r} \right \rceil$, then the reducer node has access to all the batches of files. Hence, there is no need for communication with the other nodes. So, we only consider the case when $\alpha < \left \lceil \frac{\Lambda}{r} \right \rceil$. For this case, each reducer node has access to $\alpha r$ consecutive batches of files since the content in any consecutive $\alpha $ mapper nodes are disjoint from one another. That is, for each reducer node $\lambda \in [0,\Lambda)$, the set of all batches accessible to it is $\{B_{(r\lambda+j) \text{ mod } \Lambda}: j \in [0,r\alpha)\}$. 
\begin{remark}
	The association of reducer nodes to mapper nodes in NNC-MRG is equivalent to the cyclic wrap-around model considered for multi-access coded caching problem in \cite{nihkil2017multiaccess}.
\end{remark}
 We prove that there exists a set of $l$-cyclic $g-$regular PDAs which represents MADC models with NNC-MRG in Theorem \ref{thm3} and provide a coding scheme for that model. The proof of Theorem \ref{thm3} is provided in Appendix \ref{proof thm3}.
 
\begin{thm}
	\label{thm3}
	 Consider the  $r$-cyclic $\frac{2\Lambda}{\Lambda-(\alpha -1)r}$-regular $\left( \Lambda,\Lambda,\alpha r, \frac{(\Lambda-\alpha r)(\Lambda-(\alpha-1)r)}{2}\right)$ PDAs obtained by Algorithm 2 in \cite{SR}, for some positive integers $\Lambda, r, $ and $\alpha$ such that $\alpha < \left \lceil \frac{\Lambda}{r} \right \rceil$, and $r|\Lambda$. This set of PDAs corresponds to MADC models with NNC-MRG  having $\Lambda$ batches, mapper nodes, and reducer nodes. For this model the computation load is $r$ and communication load achievable is 
	\begin{align}
	\label{L for NNC-MRG}
	L(r,\alpha) = \frac{(\Lambda-\alpha r)(\Lambda-(\alpha -1)r)}{\Lambda(\Lambda+(\alpha  -1)r)}.
	\end{align}
\end{thm}
In Theorem \ref{thm3}, we consider PDAs. The number of stars in each column is same and it is equal to $\alpha r$. Each reducer node is connected to $\alpha = \frac{Z}{r}$ mapper nodes. The communication load is a function of the parameters $\alpha$ and $r$. As $\alpha$ or $r$ changes the communication load varies. 

 Next, we illustrate Theorem \ref{thm3} via Example \ref{exmp2}.
\begin{example}
	\label{exmp2}
		We define a matrix $P_3$ as in (\ref{NNC-MRG pda eg}).
	It can be verified that $P_3$ is a $2$-cyclic $4$-regular $(12,12,8,12)$ PDA. 
	The row index $\lambda \in [0,12)$ represents the batch $B_{\lambda}$ and the column index $\lambda \in [0,12)$ represents the reducer node $\lambda$.  
	
	Consider an MADC model with NNC-MRG where there are $\Lambda=12$ mapper and reducer nodes with each reducer node connected to $\alpha =4$ neighboring mapper nodes. Let there be $N = 12$ input files $\{w_n:n\in [0,12)\}$, and $Q = 12$ output functions $\{\phi_q: q\in [0,12)\}$ to be computed. Let the  computation load be $r = 2$. We partition $12$ files into $F = 12$ batches $\{B_{\lambda}:\lambda \in [0,12)\}$, where $B_{\lambda} =\{w_{\lambda}\} $ for $ \lambda \in [0,12)$. We assign $Q/\Lambda = 1$ output function to each reducer node $\lambda \in [0,12)$. Let the indices of the output functions assigned to the reducer node $\lambda$ be $\mathcal{W}_{\lambda}= \{\lambda\}.$
	Assign a set of $2$ batches of files for each mapper node $\lambda \in [0, 12)$ as follows.
	\begin{align}
	M_0 =&M_6 = \{B_{0} , B_{1}\} , &
	M_1 =&M_7= \{B_{2} , B_{3} \}, \nonumber\\
	M_2 =&M_8= \{B_{4} , B_{5}\} , &
	M_3 =&M_9= \{B_{6} , B_{7}\} , \nonumber \\
	M_4 =&M_{10}= \{B_{8} , B_{9}\} , &
	M_5 =&M_{11}= \{B_{10} , B_{11}\} , \nonumber \\
	\end{align}
For each $\lambda \in [0,12)$, the mapper node $\lambda$ computes $Q = 12$ intermediate values for each assigned input file.
			\begin{strip}
		\begin{center}
			\begin{equation}
			P_3=\begin{blockarray}{ccccccccccccc}
			&\{0\} & \{1\}&\{2\}&\{3\}&\{4\}&\{5\} & \{6\} & \{7\}&\{8\}&\{9\}&\{10\}&\{11\} \\
			\begin{block}{c(cccccccccccc)}
			B_0&* & 0&1&*&*&* & * & 6&7&*&*&* \\
			B_1&* & 3&4&*&*&* &  * & 9&10&*&*&*\\
			B_2&* & *&2&0&*&* & 	 * & *&8&6&*&* \\
			B_3&* & *&5&3&*&* & 	 * & *&11&9&*&* \\
			B_4&* & *&*&1&2&* & 	 * & *&*&7&8&* \\
			B_5&* & *&*&4&5&* & 	 * & *&*&10&11&* \\ 
			B_6&* & *&*&*&0&1 & 	 * & *&*&*&6&7 \\
			B_7&* & *&*&*&3&4 &	 * & *&*&*&9&10 \\
			B_8&0 & *&*&*&*&2 & 	 6 & *&*&*&*&8 \\
			B_9&3 & *&*&*&*&5 &	 9 & *&*&*&*&11 \\
			B_{10}&1 & 2&*&*&*&* &	 7 & 8&*&*&*&* \\
			B_{11}&4 & 5&*&*&*&* & 	 10 & 11&*&*&*&* \\
			\end{block}
			\end{blockarray}
			\label{NNC-MRG pda eg}
			\end{equation}
		\end{center}
	\end{strip}	
	 In MADC model with NNC-MRG, each reducer node is connected to $4$ neighbouring mapper nodes in a cyclic wrap around way. That is, each reducer node $\lambda \in [0,12)$  is connected to mapper nodes $\{\lambda, (\lambda+1) \text{ mod } 12, (\lambda+2) \text{ mod } 12, (\lambda+3) \text{ mod } 12\}$. It can be observed that the array $P_3$ corresponds to this MADC model since the set of all batches assigned to each reducer node $\lambda$ is as follows:
	\begin{align}
			R_{\lambda}  =  M_{\lambda} \cup M_{(\lambda+1) \text{ mod } 12} \cup  M_{(\lambda+2) \text{ mod } 12 }\cup M_{(\lambda+3) \text{ mod } 12}.
	\end{align}
	Since $Z=8$ for the PDA $P_3$, each reducer node $\lambda\in[0,12)$ has access to $8$ batches of files and can retrieve the IVs computed from the files in those $8$ batches, i.e, it can retrieve all the IVs in $V_{\lambda} = \{v_{q,n} : q \in [0,12), w_n \in B_{f} , B_{f}\in R_{\lambda},f \in [0,12)\}$.  
	
	Let $S_{\lambda}$ denote the set of all integers present in column $\lambda$, for  $\lambda\in [0,12)$.
	Consider the first column, i.e. column with index $0$ of $P_3$. The set of all integers present in this column is $S_{0} = \{0,3,1,4\}$. We concatenate the IVs for the output functions in $\mathcal{W}_{0}$ which need to be computed by the reducer node $0$ and can be computed from the files in $B_{8}$, i.e., $\{v_{q,n} : q \in \mathcal{W}_{0}, w_n \in B_{8} \}$, into a symbol
	\begin{align}
		\U_{\W_{0},B_{8}} = (v_{q,n} : q \in \{0\}, w_n \in \{w_8\}).
	\end{align}
	Similarly, we concatenate the IVs for the output functions in $W_{0}$ which need to be computed by the reducer node $0$ and can be computed from the files in $B_{9},B_{10},$ and $B_{11}$ into the symbols $\U_{\W_{0},B_{9}},\U_{\W_{0},B_{10}},$  and $\U_{\W_{0},B_{11}} $ respectively such that 
	\begin{align}
		\U_{\W_{0},B_{9}} &= (v_{q,n} : q \in \{0\}, w_n \in \{w_9\}), \nonumber\\
		\U_{\W_{0},B_{10}} &= (v_{q,n} : q \in \{0\}, w_n \in \{w_{10}\}), \nonumber\\
		\U_{\W_{0},B_{11}} &= (v_{q,n} : q \in \{0\}, w_n \in \{w_{11}\}).
	\end{align}	
	Consider the entry $s= 0 $ in $ S_{0}$. The other entries which are $0
	$ are in the columns indexed by $1,3,$ and $4$. Hence, we partition the symbols in $\U_{\W_{0},B_{8}}$ into $3$ packets, each of equal size, 
	\begin{align}
		\U_{\W_{0},B_{8}} =\{\U_{\W_{0},B_{8}}^{1}, \U_{\W_{0},B_{8}}^{3}, \U_{\W_{0},B_{8}}^{4} \} .
	\end{align}
	Next for the entry $s= 3,1,$ and $4$ in $ S_{0}$, we partition $\U_{\W_{0},B_{9}},\U_{\W_{0},B_{10}},$  and $\U_{\W_{0},B_{11}} $ respectively into $3$ packets as follows:
	\begin{align}
		\U_{\W_{0},B_{9}} &=\{\U_{\W_{0},B_{9}}^{1}, \U_{\W_{0},B_{9}}^{3}, \U_{\W_{0},B_{9}}^{4} \}, \nonumber\\
		\U_{\W_{0},B_{10}} &=\{\U_{\W_{0},B_{10}}^{2}, \U_{\W_{0},B_{10}}^{3}, \U_{\W_{0},B_{10}}^{5} \}, \nonumber\\
		\U_{\W_{0},B_{11}} &=\{\U_{\W_{0},B_{11}}^{2}, \U_{\W_{0},B_{11}}^{3}, \U_{\W_{0},B_{11}}^{5} \}
	\end{align}
	Similarly, for each column $\lambda \in [1,12)$, we concatenate the IVs for the output functions in $\W_{\lambda}$, which need to be computed by the reducer node $\lambda$ and can be computed from the files not accessible to them into a symbol and we partition them in into $3$ packets of equal sizes.
	Since $|S_\lambda|=4, \forall \lambda \in [0,12)$, each reducer node $\lambda$ transmits $4$ coded symbols $X_s^{\lambda}$, for $s\in S_{{\lambda}}$. 
	The coded symbols transmitted by the reducer nodes are
\begin{align} 
X_s^{\lambda} = \bigoplus_{\substack{\text{$(u,v)\in [0,12) \times ([0,12) \backslash k) :$} \\ \text{$p_{u,v}=s$}}} 
\U_{\mathcal{W}_v,B_u}^{{\lambda}},  \forall {\lambda} \in[0,12), s \in S_{\lambda}.
\end{align}
	It can be verified that all the reducer nodes can retrieve all required symbols needed to compute the respective functions. A total of $48$ coded symbols are transmitted across the reducer nodes each of size $\frac{t}{3}$ bits. Hence, the communication load is $L(2,4) =\frac{\frac{t}{3} *48}{12*12*t} =\frac{1}{9}$.
\end{example}
\section{Generalized Combinatorial-MRG}
\label{GC-MRG}
In this section, we consider a generalization of MADC models with CT, which we refer as MADC models with Generalized Combinatorial-MRG (GC-MRG). First, we obtain a set of $g-$regular MRAs using {\textbf{Algorithm \ref{algo1}}}. We prove that this set of MRAs represents MADC models with CT in Theorem \ref{thm4}. We obtain an extended  set of MRAs  in {\textbf{Algorithm \ref{algo2}}} using {\textbf{Algorithm \ref{algo1}}}. We show that the set of MRAs obtained using {\textbf{Algorithm \ref{algo2}}} represents the MADC models with GC-MRG in Theorem \ref{thm5}.   We also obtain a lower bound on the communication load for MADC models with GC-MRG considering only homogeneous networks in Theorem \ref{thm6}. The proofs of correctness of {\textbf{Algorithm \ref{algo1}}}, and {\textbf{Algorithm \ref{algo2}}} are  provided in Appendices \ref{proof algo1}, and \ref{proof algo2} respectively,
while the proofs of Theorems \ref{thm4}, \ref{thm5}, and \ref{thm6} are provided in Appendices \ref{proof thm4}, \ref{proof thm5}, and  \ref{proof thm6} respectively. 
\begin{defn}
	(\textbf{Generalized Combinatorial-MRG}): Consider an MADC model with $\Lambda$ mapper nodes and $K'$ reducer nodes. An MRG is said to be a Generalized Combinatorial-MRG (GC-MRG), if it satisfies the following conditions:
	\begin{enumerate}
		\item There are $F={\Lambda \choose r}$ batches, $B_{T}$ with $ T \subset [0,\Lambda)$ and $ |T|=r$, each containing $\eta_1 = N / F$ files, for some $r \in [\Lambda]$.
		\item Each mapper node $\lambda \in [0,\Lambda)$ can access a batch $B_{T}$ if $\lambda \in T$, that implies $
		M_{\lambda} = \{B_{T} : T \subset [0, \Lambda), |T|=r,\lambda \in T\}.$
		\item Every combination of $\alpha$  mapper nodes is uniquely connected to $K_{\alpha} \geq 0$ reducer nodes, for each $\alpha \in [\Lambda]$, i.e., $K' = \sum_{\alpha \in [\Lambda]} K_{\alpha} {\Lambda \choose \alpha}$
	\end{enumerate}
\end{defn}
\begin{remark}
	The MADC models with CT defined in \cite{BP} is a GC-MRG with fixed $\alpha \in [\Lambda]$ and with $K_{\alpha} =1$. Hence the total number of reducer nodes reduces to ${\Lambda \choose \alpha}$, i.e., for every combination of $\alpha$ mapper nodes, there is a unique reducer node which those mapper nodes are connected to.
\end{remark}
For MADC models with GC-MRG, we assume that in the map phase the input database is split into $F={\Lambda \choose r}$ disjoint batches $B_{T}$ with $ T \subset [0,\Lambda)$ and $ |T|=r$, each containing $\eta_1 = N / F$ files, for some $r \in [\Lambda]$. Consequently, we have a batch of files for each $T \subset [0, \Lambda)$ such that $|T|=r,$ which implies $\bigcup_{T \subset [0, \Lambda): |T|=r} B_{T} = \{w_0,w_1,\ldots,w_{N-1}\}$. The mapper node $\lambda \in [0,\Lambda)$ is assigned all batches $B_{T}$ if $\lambda \in T$, that implies $
M_{\lambda} = \{B_{T} : T \subset [0, \Lambda), |T|=r,\lambda \in T\}.$ This phase is fixed.  For every combination of $\alpha$ mapper nodes, $\alpha\in [\Lambda]$,  there are $K_{\alpha}$ reducer nodes to which those mapper nodes are uniquely connected to.

Divide $K'$ reducer nodes into $\Lambda$ blocks, $\{A_{\alpha} :\alpha \in [\Lambda]\}$, such that all the reducer nodes connected to exactly $\alpha$ mapper nodes are put in the block $A_{\alpha}$, for each $\alpha \in [\Lambda]$. Each block $A_{\alpha}$ contains $K_{\alpha}{\Lambda \choose \alpha}$ reducer nodes. If $\alpha > \Lambda-r$, then the reducer nodes in $A_{\alpha}$ have access to all the batches of files,  since every batch $B_{T}$, for $T\subset [0,\Lambda)$ and $|T|=r$, is mapped to $r$ mapper nodes.  Hence, there is no need for exchanging the IVs. We will be considering only the cases when $\alpha \leq [\Lambda -r]$, i.e., we are interested in the set of reducer nodes in the set $\cup_{\alpha \in [\Lambda-r]} A_{\alpha}$. Hence, the effective number of reducer nodes that we will be considering for GC-MRG is $K= \sum_{\alpha \in [\Lambda-r]} K_{\alpha} {\Lambda \choose \alpha}$.
\begin{thm}
	\label{thm4}
	Consider a set of $g-$regular MRAs 
		\begin{align}
		\Biggl \{g\text{-regular }\left ({\Lambda \choose \alpha}, {\Lambda \choose r},  {\Lambda \choose  \alpha +r} \right ) \nonumber \text{ MRA}: \\ r \in [\Lambda-\alpha], g= {r+\alpha \choose r} \Biggr \}
		\end{align} 
	constructed using {\textbf{Algorithm \ref{algo1}}}, for some positive integers $\Lambda$ and $\alpha$ such that $ \alpha \in [\Lambda-1]$ . This set of MRAs
	corresponds to MADC models with CT having $\Lambda$ mapper nodes and $K = {\Lambda \choose \alpha}$ reducer nodes such that each reducer node is connected to distinct  $\alpha$ mapper nodes, achieving a communication load of
		\begin{align}
		L(r,\alpha) = \frac{{\Lambda -\alpha\choose r}}{{\Lambda \choose r} \left ( {r+\alpha \choose r} -1\right )}, &&&\forall r \in [\Lambda-\alpha].
		\label{L for Combinatorial}
		\end{align}
\end{thm}
In CT, the number of mapper nodes to which each reducer node is connected to  is fixed and is given by the term $\alpha$. The MRA considered in Theorem \ref{thm4} is a function of  $\alpha$ and $r$. Hence, the communication load also varies with $\alpha$ and $r$ and is a function of both the variables.
\begin{algorithm}
	\caption{${r+\alpha \choose r}\text{-regular }\left ({\Lambda \choose \alpha}, {\Lambda \choose r}, {\Lambda \choose  \alpha +r} \right ) $ MRA construction, for some positive integers $\Lambda,r$ and $\alpha$ such that $ \alpha \in [\Lambda-1]$ and $r \in [\Lambda-\alpha] $.}
	\label{algo1}
	\begin{algorithmic}[1]
		\Procedure{{\bf 1}: }{}
		Arrange all subsets of size $\alpha +r$ from $[0,\Lambda)$ in lexicographic order and for any subset $T'$ of size $\alpha+r$, define $y_{\alpha+r}(T')$ to be its order minus 1.
		\EndProcedure \textbf{ 1}
		\Procedure{{\bf 2}: Obtain an array  $D_{\Lambda,r,\alpha}$ of size ${\Lambda \choose r} \times {\Lambda \choose \alpha}$.}{}
		Denote the rows by the sets in $\{T\subset [0,\Lambda), |T| = r \}$ and columns by the sets in $\{U \subset [0,\Lambda) : |U| = \alpha\}$. Define each entry $d_{T,U}$ corresponding to the row $T$ and the column $U$, as
		\begin{align}
		\label{Dk}
		d_{T,U} = \left\{
		\begin{array}{cc}
		*, &  \text{if } |T \cap U| \neq 0 \\
		y_{\alpha+r}(T \cup U), &  \text{if } |T \cap U| = 0 
		\end{array} \right\}.
		\end{align}
		\EndProcedure \textbf{ 2}
	\end{algorithmic}
\end{algorithm}
\begin{algorithm}
	\caption{$\left (\left ( \sum_{\alpha \in [\Lambda-r]}K_{\alpha} {\Lambda \choose \alpha} \right ), {\Lambda \choose r}, \sum_{\alpha\in [\Lambda-r]} K_{\alpha}{\Lambda \choose  \alpha +r} \right ) $ MRA construction,  for some positive integers $\Lambda,r, $ and $K_{\alpha}$ such that $r \in [\Lambda-1]$.}
	\label{algo2}
	\begin{algorithmic}[1]
		\Procedure{{\bf 1}: }{}
		\For{\textit{$\alpha \in \left [\Lambda-r \right ]$}}
		\State	 Construct an ${\Lambda \choose r} \times {\Lambda \choose \alpha}$ array $D_{\alpha,1}$ using 
		\State\textbf{Algorithm \ref{algo1}}.
		\EndFor
		\EndProcedure \textbf{ 1}
		\Procedure{{\bf 2}: }{}
		\For{\textit{$\alpha \in \left [\Lambda-r \right ]$}}
		\State Create an ${\Lambda \choose r} \times K_{\alpha} {\Lambda \choose \alpha}$ array 
		\State $\textbf{D}_{\alpha} =  \left ( \begin{array}{cccccc}
		{{D}}_{\alpha,1} & {{D}}_{\alpha,2} & {{D}}_{\alpha,3} &\ldots &{{D}}_{\alpha, K_{\alpha}}
		\end{array}\right )$
		\State where  every  array $ {{D}}_{{\alpha,m}} ={{D}}_{\alpha,1}+(m-1){S}_1,$ for 
		\State $m \in  [2,K_{\alpha}]$ and $  {S}_1={\Lambda \choose \alpha + r},$ is a ${\Lambda \choose r} \times {\Lambda \choose \alpha}$ array,
		\State with $ *+(m-1){S}_1 =*$. 
		\EndFor
		\EndProcedure \textbf{ 2}
		\Procedure{{\bf 3}: }{}
		Construct  an ${\Lambda \choose r} \times \left ( \sum_{\alpha \in [\Lambda-r]}K_{\alpha} {\Lambda \choose \alpha} \right )$ array 
		$$
		\textbf{D}=  \left ( \begin{array}{cccccc}
		\tilde{{\textbf{D}}}_1 & \tilde{\textbf{D}}_2 & \tilde{\textbf{D}}_3 &\ldots &\tilde{\textbf{D}}_{\Lambda-r}
		\end{array}\right )$$
		where  $\tilde{{\textbf{D}}}_1 = {{\textbf{D}}}_1$ and every other array $ \tilde{\textbf{D}}_{\alpha} ={\textbf{D}}_{\alpha}+S'_{\alpha},$ for $\alpha \in [2,\Lambda-r]$ and $S'_{\alpha} = \sum_{m\in [\alpha-1]}K_{m}{\Lambda \choose m + r},$ is a ${\Lambda \choose r} \times K_{\alpha}{\Lambda \choose \alpha}$ array, with $ *+S'_{\alpha} =*$. 
		\EndProcedure \textbf{ 3}
	\end{algorithmic}
\end{algorithm}
\begin{thm}
	\label{thm5}
	Consider a set of  MRAs 
	{\small
		\begin{align}
		\Biggr \{\left ( \sum_{\alpha \in [\Lambda-r]}K_{\alpha} {\Lambda \choose \alpha} , {\Lambda \choose r}, \sum_{\alpha\in [\Lambda-r]} K_{\alpha}{\Lambda \choose  \alpha +r} \right ) \text{ MRA}: \nonumber\\  r \in [\Lambda-1] \Biggr \}
		\end{align} }
	
	\noindent constructed using {\textbf{Algorithm \ref{algo2}}},  for some positive integers $\Lambda $ and $K_{\alpha}$. This set of MRAs corresponds to MADC models with GC-MRG having $\Lambda$ mapper nodes and $K = \sum_{\alpha \in [\Lambda-r] } K_{\alpha}{\Lambda \choose \alpha}$ reducer nodes such that every combination of $\alpha$  mapper nodes is uniquely connected to $K_{\alpha}$ reducer nodes, for each $\alpha \in [\Lambda-r]$. The computation load for this MADC model is $r$ and  communication  load achievable is given by
	\begin{align}
	L(r) = \frac{1}{K}\sum_{\alpha\in [\Lambda-r]}\frac{ K_{\alpha}{\Lambda-r \choose \alpha }}{ \left({r+\alpha \choose r}-1\right)}, \forall r \in [\Lambda-1].
	\label{L for GC-MRG}
	\end{align}
\end{thm}
In GC-MRG, the number of mapper nodes to which a reducer node is connected to is not fixed. It varies from $1$ to $\Lambda-r$ since we consider all the possible combinations in GC-MRG.
\begin{thm}
	\label{thm6}
	A lower bound on the computation-communication load curve for an MADC model with GC-MRG for homogeneous networks that consists of $\Lambda$ mapper nodes,  and $K = \sum_{\alpha \in [\Lambda-r] } K_{\alpha}{\Lambda \choose \alpha}$ reducer nodes, is given by the lower convex envelope of the points 
	\begin{align}
		\label{improved lb}
		\left (r,L_{new}^{lb}(r)\right) =  \Biggl  \{\left ( r, \frac{ \sum_{\alpha\in [\Lambda-r]}K_{\alpha}{\Lambda-r \choose \alpha} }{K\sum_{\alpha\in [\Lambda-r]}K_{\alpha}\left ({\Lambda \choose \alpha} -{\Lambda - r \choose \alpha} \right )}\right ): \nonumber \\ r \in [\Lambda-1] \Biggr \} .
	\end{align}
\end{thm}
\begin{cor}
	 For a fixed $\alpha \in [\Lambda]$ and $K_{\alpha}=1$, the GC-MRG reduces to CT. For MADC models with CT, when $\alpha =1$, each reducer node is assigned exactly one unique mapper node which corresponds to the original DC  model \cite{LMA}. Hence, we have $K=\Lambda$. For this setting, our lower bound in Theorem \ref{thm5} coincides with the lower bound provided for the DC problem, with $K$ servers and computation load $r$ in \cite{LMA}, which is given by $L_{new}^{lb} (r)= \frac{\Lambda- r}{\Lambda r}$.
	In \cite{BP}, for $\alpha =1$, the lower bound is given by $L_{BE}^{lb} = \frac{\Lambda -r}{\Lambda (r+1)}$, which does not coincide with the lower bound for the DC model. 
\end{cor}
\begin{cor}
	For MADC models with CT, if $\alpha = \Lambda -r$, then the lower bound proposed in Theorem (\ref{thm6}) matches the achievable communication load given in Theorem \ref{thm4}, which is given by 
	\begin{equation}
	L(r, \alpha) =L_{BE}(r) = L_{new}^{lb} (r) =\frac{1}{{r+\alpha \choose r} \left ({r+\alpha \choose r} -1\right )}
	\end{equation}
	Hence, the BE scheme is optimal for an MADC model with CT if $\alpha = \Lambda - r$. 
\end{cor}
Note that the lower bound derived in \cite{BP} for MADC models with CT is for general heterogeneous networks while we consider only homogeneous setting.
\begin{remark}
	From the proof of correctness of {\bf Algorithm \ref{algo1}} provided in Section \ref{proof algo1}, we notice that the symbol $*$ appears exactly  $Z={\Lambda \choose r} - {\Lambda -\alpha \choose r}$ times in each column. Hence, the MRA is a ${r+\alpha \choose r}\text{-regular }\left ({\Lambda \choose \alpha}, {\Lambda \choose r}, {\Lambda \choose r} - {\Lambda -\alpha \choose r},{\Lambda \choose  \alpha +r} \right ) $ PDA. We observe that when $\alpha =1$, the MRA that we obtain using {\bf Algorithm \ref{algo1}} is equivalent to the PDA corresponding to the coded caching schemes in \cite{MAN,YCTCPDA}. Also, when $\alpha>1$, the MRA obtained is equivalent to the PDA corresponding to the multi-access coded caching schemes in \cite{MKR}. 
\end{remark}
\noindent Now, we illustrate the BE scheme by using MRAs through an example.
\begin{example}
	\label{exmp3}
	Consider Example \ref{exmp6}, i.e, an MADC model with CT with $\Lambda = 4$ mapper nodes and $K = {\Lambda \choose \alpha}= 6$ reducer nodes, where $\alpha = 2$.
	 We define an array $P_4$, as follows.
		\begin{equation}
		\begin{blockarray}{ccccccc}
		& \{01\} & \{02\} & \{03\} &   \{12\} & \{13\} & \{23\}  \\
		\begin{block}{c(cccccc)}
		B_{\{01\}} & * & * & * & * & * & 0  \\
		B_{\{02\}} & * & * & * & * & 0 & *  \\
		B_{\{03\}} & * & * & * & 0 & * & * \\
		B_{\{12\}} & * & * & 0 & * & * & *  \\
		B_{\{13\}} & * & 0 & * & * & * & *  \\
		B_{\{23\}} & 0 & * & * & * & * & *  \\
		\end{block}
		\end{blockarray}
		\end{equation}
	It can be easily verified that $P_4$ is a $6-$regular $(6, 6,  1)$ MRA. The rows represent the blocks $\{B_T : T \subset [0,4), |T|=2\}$ and the columns represent the reducer nodes $\{U: U \subset [0,4), |U|=2\}$.
	From the array $P_4$, the set of all files accessible to each reducer node $U$ is as follows:
	\begin{align}
		R_{\{01\}}  =\{B_{\{01\}},B_{\{02\}},B_{\{03\}},B_{\{12\}},B_{\{13\}}\}  \nonumber\\
		R_{\{02\}}  =\{B_{\{01\}},B_{\{02\}},B_{\{03\}},B_{\{12\}},B_{\{23\}}\}  \nonumber\\
		R_{\{03\}}  =\{B_{\{01\}},B_{\{02\}},B_{\{03\}},B_{\{13\}},B_{\{23\}}\} \nonumber\\
		R_{\{12\}}  =\{B_{\{01\}},B_{\{02\}},B_{\{12\}},B_{\{13\}},B_{\{23\}}\} \nonumber\\
		R_{\{13\}}  =\{B_{\{01\}},B_{\{03\}},B_{\{12\}},B_{\{13\}},B_{\{23\}}\} \nonumber\\
		R_{\{23\}}  =\{B_{\{02\}},B_{\{03\}},B_{\{12\}},B_{\{13\}},B_{\{23\}}\} 
	\end{align}
	and each reducer node $U$ can retrieve all IVs in $V_U = \{v_{q,n} : q \in [0,6), w_n  \in B_{T}, B_{T}\in R_U\}$.
	Consider the first column, i.e. column with index $U=\{01\}$ of $P_4$. The set of all integers present in this column is $S_{\{01\}} = \{0\}$. We concatenate the IVs for the output functions in $\W_{\{01\}}$ which needs to be computed by the reducer node $\{01\}$ and can be computed from the files in $B_{\{23\}}$, i.e., $\{v_{q,n} : q \in \W_{\{01\}}, w_n \in B_{\{23\}} \}$, into a symbol
	\begin{align}
		\U_{\W_{\{01\}},B_{\{23\}}} = (v_{q,n} : q \in \{0\}, w_n \in \{w_5\}).
	\end{align}
	For the entry $s= 0 \in S_{\{01\}}$, the other entries which are $0$  are in the columns $\{02\},\{03\},\{12\} ,\{13\}$ and $\{23\}$. Hence, we partition the symbols in $\U_{\W_{\{01\}},B_{\{23\}}}$ into $(g-1) = 5$ packets, each with equal size such that
	\begin{align}
		\U_{\W_{\{01\}},B_{\{23\}}} =\biggl \{&\U_{\W_{\{01\}},B_{\{23\}}}^{\{02\}}, \U_{\W_{\{01\}},B_{\{23\}}}^{\{03\}},\U_{\W_{\{01\}},B_{\{23\}}}^{\{12\}}, \nonumber\\& \U_{\W_{\{01\}},B_{\{23\}}}^{\{13\}},\U_{\W_{\{01\}},B_{\{23\}}}^{\{23\}} \biggr \}.
	\end{align}
	Similarly, for each column $U$, we concatenate the IVs for the output functions in $W_{U}$ which needs to be computed by the reducer node $U$ and can be computed from the files not accessible to them, 
	and partition them into $5$ packets of equal sizes. 
	Each reducer node $U$ transmits one coded symbol $X_0^U$. The  coded symbols transmitted by the reducer nodes are as (\ref{trans}).
	
	\begin{strip}
		\begin{center}
			\begin{equation}
			\label{D}
			\begin{blockarray}{ccccccccccccccccccccccccccc}
			&&& &\multicolumn{2}{c}{${\bf D}_{1}$}&& && &&&&&& &\multicolumn{4}{c}{${{\bf D}_{2}}+8$}&&&&&&\\
			\begin{block}{c(cccccccc|cccccccccccccccccc)}
			B_{01}&*&*&0&1&*&*&4&5&*&*&*&*&*&8&*&*&*&*&*&9&*&*&*&*&*&10\\
			B_{02}&*&0&*&2&*&4&*&6&*&*&*&*&8&*&*&*&*&*&9&*&*&*&*&*&10&*\\
			B_{03}&*&1&2&*&*&5&6&*&*&*&*&8&*&*&*&*&*&9&*&*&*&*&*&10&*&*\\
			B_{12}&0&*&*&3&4&*&*&7&*&*&8&*&*&*&*&*&9&*&*&*&*&*&10&*&*&*\\
			B_{13}&1&*&3&*&5&*&7&*&*&8&*&*&*&*&*&9&*&*&*&*&*&10&*&*&*&*\\
			B_{23}&2&3&*&*&6&7&*&*&8&*&*&*&*&*&9&*&*&*&*&*&10&*&*&*&*&*\\
			\end{block}
			\end{blockarray}
			\end{equation}
		\end{center}	
	\end{strip}	
	As stated in Example \ref{exmp6}, it can be verified that all nodes can retrieve all required symbols needed to compute the respective functions. The communication load for this example is $L(2,2) =\frac{1}{30}$.
 The lower bound in Theorem (\ref{thm3}) is equal to $L_{new}^{lb}(2) = \frac{1}{6*5}=\frac{1}{30}$. Hence, for this example BE scheme is optimal under CT. 
\end{example}
\begin{example}
	\label{GC-MRG eg}
	Consider the array $P_5$ in (\ref{D}). It can be verified that $P_5$ is an $(26, 6,  11)$ MRA.  The MRA  $P_5$ is obtained from {\bf Algorithm \ref{algo2}}. We take $\Lambda=4,r=2, K_1=2,$ and $K_2=3$. For $\alpha=1,$ and $2$, using {\bf procedure 1}, we obtain two arrays $D_{1,1}$ and $D_{2,1}$ respectively as follows.
	\begin{figure}[H]
		\begin{subfigure}{.22\textwidth}
				\begin{equation}
				D_{1,1}=	\begin{blockarray}{cccc}
					\begin{block}{(cccc)}
						* & * & 0 & 1  \\
						* & 0 & * & 2  \\
						* & 1 & 2 & *  \\
						0 & * & * & 3   \\
						1 & * & 3 & *  \\
						2 & 3 & * & *   \\
					\end{block}
				\end{blockarray} \nonumber
			\end{equation} 
		\end{subfigure}
		\begin{subfigure}{.2\textwidth}
				\begin{equation}
				D_{2,1}=	\begin{blockarray}{cccccc}
					\begin{block}{(cccccc)}
						* & * & * & * & * & 0  \\
						* & * & * & * & 0 & *  \\
						* & * & * & 0 & * & * \\
						* & * & 0 & * & * & *  \\
						* & 0 & * & * & * & *  \\
						0 & * & * & * & * & *  \\
					\end{block}
				\end{blockarray} \nonumber
			\end{equation} 
		\end{subfigure} 
	\end{figure}
	From $D_{1,1}$ and $D_{2,1}$, using {\bf procedure 2}, we obtain the arrays ${\bf D}_{1}$ and ${\bf D}_{2}$ respectively as in (\ref{D1}) and (\ref{D2}).
		\begin{equation}
			\label{D1}
			{\bf D}_{1} =\begin{blockarray}{cccccccc}
				& \multicolumn{2}{c}{${ D}_{1,1}$}&&& \multicolumn{2}{c}{${ D}_{1,1}+4$}&\\
				\begin{block}{(cccc|cccc)}
					*&*&0&1&*&*&4&5\\
					*&0&*&2&*&4&*&6\\
					*&1&2&*&*&5&6&*\\
					0&*&*&3&4&*&*&7\\
					1&*&3&*&5&*&7&*\\
					2&3&*&*&6&7&*&*\\
				\end{block}
			\end{blockarray} 
		\end{equation}

	\begin{figure*}
	\begin{center}
		\begin{equation}
			\label{D2}
			{\bf D}_{2}=	\begin{blockarray}{cccccccccccccccccc}
				& \multicolumn{4}{c}{${{ D}_{2,1}}$}&&& \multicolumn{4}{c}{${{ D}_{2,1}}+1$}&&&\multicolumn{4}{c}{${{ D}_{2,1}}+2$}&\\
				\begin{block}{(cccccc|cccccc|cccccc)}
					*&*&*&*&*&0&*&*&*&*&*&1&*&*&*&*&*&2\\
					*&*&*&*&0&*&*&*&*&*&1&*&*&*&*&*&2&*\\
					*&*&*&0&*&*&*&*&*&1&*&*&*&*&*&2&*&*\\
					*&*&0&*&*&*&*&*&1&*&*&*&*&*&2&*&*&*\\
					*&0&*&*&*&*&*&1&*&*&*&*&*&2&*&*&*&*\\
					0&*&*&*&*&*&1&*&*&*&*&*&2&*&*&*&*&*\\
				\end{block}
			\end{blockarray}
		\end{equation}
	\end{center}	
\end{figure*}
Finally, we obtain the array $P_5= {\bf D}$ as in (\ref{D}) using {\bf procedure 3}.
\end{example}
\noindent Let us take an example for MADC models with GC-MRG.
\begin{example}
	\label{exmp5}
	Consider an MADC model with GC-MRG with $\Lambda = 4$ mapper nodes.
	Assume that we have $N = 6$
	input files $\{w_0,w_1,w_2,w_3,w_4,w_5\}$. 	We partition $N = 6$ files into ${\Lambda \choose r} =6$ disjoint batches $B_{T}: T \in \{01,02,03,12,13,23\},$ where $r=2,$ i.e., we have
	\begin{align}
		B_{\{01\}} &= \{w_0\},&
		B_{\{02\}} &= \{w_1\},&
		B_{\{03\}} &= \{w_2\}, \nonumber\\
		B_{\{12\}} &= \{w_3\},&
		B_{\{13\}} &= \{w_4\},&
		B_{\{23\}} &= \{w_5\}.
	\end{align}
	For each $\lambda \in [0,4)$, mapper node $\lambda \in [0, 4)$ is assigned the set of files in $B_{T}$ if $\lambda \in T$, i.e., we have
	\begin{align}
		M_0 &= \{B_{\{01\}} , B_{\{02\}} , B_{\{03\}}\}, \nonumber\\
		M_1 &= \{B_{\{01\}} , B_{\{12\}} , B_{\{13\}} \}, \nonumber\\
		M_2 &= \{B_{\{02\}} , B_{\{12\}} , B_{\{23\}} \}, \nonumber\\
		M_3 &= \{B_{\{03\}} , B_{\{13\}} , B_{\{23\}}\} .
	\end{align}
	Each mapper node $\lambda$ computes $Q = 6$ intermediate values for each assigned input file.
	
	 We take the $(26,6,11)$ MRA $P_5$ considered in Example \ref{GC-MRG eg}.
	Suppose we have $K=\sum_{\alpha \in [2]}K_{\alpha}{\Lambda \choose \alpha} =26$ reducer nodes with $K_1=2$ and $K_2=3$ and $Q = 26$ output functions to be computed across the reducer nodes. We assign $Q/K = 1$ output functions to each reducer node.
	 The rows represent the blocks $\{B_T : T \subset [0,4), |T|=2\}$ and the columns represent the reducer nodes. The first 8 columns represent the reducer  nodes which are exactly connected to $1$ mapper node ($\alpha =1$) while the rest of the columns represents the reducer nodes which are exactly connected to $2$ mapper nodes $\alpha =2$.  We also observe from Example \ref{GC-MRG eg} that $K_1=2$ and $K_2=3$. Hence, this MRA represent our MADC model.

	The communication load given by Theorem \ref{thm5} is $L(2)=0.046$ while the lower bound given by Theorem (\ref{thm6}) is equal to $L_{new}^{lb}(2) =0.014$.
\end{example}
\noindent Let us consider another example with the same number of mapper nodes, computation load and $\alpha$ as in Example \ref{exmp2}. Unlike Example \ref{exmp2}, let us use CT now.
\begin{example}
	\label{exmp4}
	Consider an MADC model with CT with $\Lambda = 12$ mapper nodes and $K = {\Lambda \choose \alpha}= 495$ reducer nodes, where $\alpha = 4$. Assume that we have $N = 66$ input files and $Q = 495$ output functions to be computed across the reducer nodes. We assign $Q/K = 1$ output functions to each reducer  node.
Assume that the computation load is $r=2$. We partition $N = 15$ files into $F = 15$ batches $\{B_T:T\subset [0,12), |T|=2\}$ and each mapper node $\lambda \in [0,12)$ is assigned a set of files in $B_T$ if $\lambda \in T$, for $T\subset [0,12)$ such that $ |T|=2$.
	The communication load achievable using BE scheme for such a model is $L(2,4) =0.3$.
\end{example}	
\begin{table}
	\begin{center}
			\begin{tabular}{ | c|c|c|}
			\hline	
			\textbf{Parameters}&  \textbf{Example \ref{exmp2}} & \textbf{Example \ref{exmp4} }\\  \hline \hline
			\textit{$\Lambda$}  &12& 12\\ \hline
			\textit{ $K$} &12&495 \\ \hline
			\textit{$r$} &2&2 \\ \hline
			\textit{$\alpha$} &4&4 \\ \hline
			\textit{$N$}  &12& 66\\ \hline
			\textit{ $Q$}  &12&495 \\ \hline
			\textit{$F$}  &12&66 \\ \hline
			\textit{$L$} &0.11& 0.03\\ \hline
		\end{tabular}
		\caption{Comparison of Examples \ref{exmp2} and \ref{exmp4}.}
		\label{tab1}
				\end{center}
	\end{table}
	\begin{table}
		\begin{center}
			\begin{tabular}{ | c|c|c|}
			\hline	
			\textbf{Parameters}&  \textbf{NNC-MRG} & \textbf{CT} \\  \hline \hline
			\textit{No. of mapper nodes: $\Lambda$}  &$\Lambda$& $\Lambda$\\ \hline
			\textit{No. of reducer nodes: $K$} &$\Lambda$&${\Lambda \choose \alpha}$ \\ \hline
			\textit{Computation load: $r$} &$r$&$r$ \\ \hline
			\textit{No. of batches: $F$}  &$\Lambda$&${\Lambda \choose r}$ \\ \hline
			\textit{No. of files: $N$}  &$\eta_1 \Lambda$& $\eta_1 {\Lambda \choose r}$\\ \hline
			\textit{No. of output functions: $Q$}  &$\eta_2 \Lambda$&$\eta_2 {\Lambda \choose \alpha}$ \\ \hline
			\textit{Communication load: $L$} &(\ref{L for NNC-MRG})& (\ref{L for Combinatorial})\\ \hline
		\end{tabular}
		\caption{Comparison of NNC-MRG and CT for fixed $\alpha$.}
		\label{tab2}
\end{center}
\end{table}

\begin{remark}
	\label{remark3}
	From Property 1, we know that an array $\hat{P}$ obtained by removing some of the columns in an 
	MRA is also an MRA as long as condition {\it C1} holds. One of the major advantages of this property is that in MADC models with GC-MRG, if we have only a subset of reducer nodes available, then the truncated MRA obtained by removing the corresponding columns from the MRA obtained using {\bf Algorithm \ref{algo2}} serves as the MRA for the corresponding problem, as long as condition {\it C1} holds. Hence, using those truncated MRAs we can have coding schemes for MADC models with GC-MRG with lesser number of reducer nodes as well.
\end{remark}
Remark \ref{remark3} specifies an important advantage of representing MADC models via MRAs. We can easily come up with coding schemes for CT even if some of the reducer nodes are absent. This is especially useful for larger values of $\Lambda$ as in Example \ref{exmp4}. The total number of reducer nodes, and output functions required in Example \ref{exmp4} are $495$.  If we have less than $495$ reducer nodes available, we cannot use the BE scheme. In particular, if $K<{\Lambda \choose \alpha}$, BE scheme cannot be used. 

Another way of dealing with the exponential increase in the number of reducer nodes and files in case of CT is by choosing a different MRG.
Comparing Examples \ref{exmp2} and \ref{exmp4} (tabulated in Table \ref{tab1}), we observe that the number of mapper nodes, computation load, and $\alpha$ are the same in both examples. The total number of reducer nodes, and output functions required in Example \ref{exmp4} are $495$, while it is $12$ in Example \ref{exmp2}.  Similarly, the total number of files required for Example \ref{exmp4} is $66$, while it is $12$ in Example \ref{exmp2}. Thus, the advantages of using NNC-MRG are two-fold: in terms of the required number of files as well as reducer nodes. This is achieved at the expense of a slight increase in the communication load.  A general comparison of NNC-MRG and CT is provided in Table \ref{tab2}.

Thus, we observe that by designing a MRA with appropriate parameters, one can have coding schemes for new MRGs which can potentially perform better than the CT in terms of flexibility in choosing values of $F$ and $K$. 

\section{conclusion}
In this paper, we have used a 2-layered bipartite graph named MRG and an array named MRA to represent MADC models. We connected MRAs to MRGs and provided a new coding scheme with the help of the MRA structure. We considered a new set of MRGs named NNC-MRGs and proved that a set of $l-$cyclic $g-$regular PDAs represents these MRGs and provided coding scheme for MADC models with NNC-MRGs.
We also considered a generalized version of CT named as GC-MRGs and generated a set of MRAs to represent  MADC models with GC-MRGs.
Exploring various classes of MRAs is an interesting future direction as it offers solutions for diverse MADC models. This approach aids in minimizing communication load during the shuffling phase, and provides enhanced flexibility regarding the number of reducer nodes and files required for the model. 
\section*{Acknowledgement}
This work was supported partly by the Science and Engineering Research Board (SERB) of Department of Science and Technology (DST), Government of India, through J.C Bose National Fellowship to Prof. B. Sundar Rajan. This research is also supported by the ZENITH Research and Leadership Career Development Fund and the ELLIIT funding endowed by the Swedish government to Prof. Onur Günlü.

\appendices
\section{Proof of Theorem \ref{thm1}}
\label{proof thm1}
We present the proof of Theorem \ref{thm1} in this section. Based on an $(K, F, S)$ MRA $P = [p_{f,k}]$ with $f \in [0, F)$
and $k \in [0, K)$, an MADC scheme for a model having  $K$ reducer nodes with each reducer node connected to some mapper nodes can be obtained as follows. We consider $Q=\eta_2 K$ output functions, for some integer $\eta_2$, so that each reducer node is assigned $\eta_2$ output functions to compute.

First, files are divided by grouping $N$ files into $F$ disjoint batches $\{B_{0},B_{1},\ldots, B_{F-1}\}$ each containing $ \eta_1 =  N / F$ files such that $\bigcup_{m=0}^{F-1} B_{m} = \{w_0,w_1,\ldots,w_{N-1}\}$. The rows in the MRA represent these $F$ batches $\{B_{0},B_{1},\ldots, B_{F-1}\}$. The columns in the MRA represent the  $K$ reducer nodes.  Each mapper node is assigned a subset of batches in such a way that each reducer node $k \in [0,K)$ can access all the batches in the set
$ \{B_{f} : p_{f,k} = *, f \in [0,F)\}.$

\subsection{Shuffle Phase}
Since each reducer node $k \in [0,K)$ can access all the batches in the set
\begin{align}
\label{cached}
R_k = \{B_{f} : p_{f,k} = *, f \in [0,F)\}
\end{align}
it can retrieve  IVs 
\begin{align}
\label{IVs}
V_{k}= \{v_{q,n} : q \in [0,Q), w_n \in B_{f},  B_{f} \in R_k, f\in [0,F)\}
\end{align}
where $V_{k}$ is the set of IVs that can be computed from the files accessible to the reducer node $k$. For each pair $(f, k) \in [0,F) \times [0, K)$ such that
$p_{f,k} = s \in [0,S)$, let $g_s$ be the number of occurrences of $s$. Assume that the other $g_s-1$ occurrences of $s$ are $
p_{f_1,k_1} = p_{f_2,k_2} = \ldots = p_{f_{g_s-1},k_{g_s-1}} = s. $
For each $k_i \in \{k_1,k_2,\ldots,k_{g_s-1}\}$ we know that $p_{f,k_i} =* $ (since $f \neq f_i$)  from C2-2. 
We concatenate the set of IVs for the output functions in $\mathcal{W}_k$ which needs to be computed by the reducer node $k$ and can be computed from the files in $B_f$, i.e., $\{v_{q,n} : q \in \mathcal{W}_k, w_n \in B_f \}$, into the symbol 
\begin{align}
\label{symbols_pda}
\U_{\mathcal{W}_k,B_f} = (v_{q,n} : q \in \mathcal{W}_k, w_n \in B_f) \in \F_{2^{\eta_1 \eta_2 t}}.
\end{align}
\noindent We partition the symbols in $\U_{\mathcal{W}_k,B_f}$ into $g_s-1$ packets each of equal size, i.e., we have
\begin{align}
\label{partition symbols}
\U_{\mathcal{W}_k,B_f}=\{\U_{\mathcal{W}_k,B_f}^{k_1}, \U_{\mathcal{W}_k,B_f}^{k_2}, \ldots, \U_{\mathcal{W}_k,B_f}^{k_{g_s-1}}\}.
\end{align}
\noindent Let $S_k$ be the set of integers in column indexed by $k$. For each entry $s \in S_k$, the reducer node $k$ creates a coded symbol
\begin{align}
\label{transmission}
X_s^k = \bigoplus_{(u,v)\in [0,\Lambda) \times ([0,K) / k) : p_{u,v}=s} \U_{\mathcal{W}_v,B_u}^{k}
\end{align}
\noindent and multicasts the sequence 
$	{\bf X}_k = \{X_k^s: s \in S_k\}.$
The reducer node $k$ can create the coded symbol $X_k^s$ from IVs accessible to it. In fact, for each $(u, v)$ in the sum (\ref{transmission}), there exists some $f \in [0, F)$ such that, $p_{u,v} = p_{f,k} = s$. Since $v$ $ \neq k$, we know that $u$ $ \neq f$ and $ p_{u,k} = *$ from C2-2. Thus, the reducer node $k$ has access to IVs $\{v_{q,n} : q \in \mathcal{W}_v, w_n \in B_u\}$ from (\ref{cached}) and (\ref{IVs}) and, thus, can create the symbol $\U_{\mathcal{W}_v,B_u}^{k}$ from (\ref{symbols_pda}) and (\ref{partition symbols}).
\subsection{Reduce Phase}

Receiving the sequences $\{{\bf X}_j\}_{j \in [0,K)\backslash k}$, each reducer node $k$ decodes all IVs of its output functions, i.e., $\{v_{q,n} : q \in \mathcal{W}_k, n \in [0,N)\}$, with the help of IVs $\{v_{q,n} : q \in [0,Q), w_n \in B_{f}, B_f\in R_k, f \in [0,F) \}$ it has access to, and finally compute the output functions assigned to it.

The reducer node $k$ can compute the output functions in $\mathcal{W}_k$ in the reduce phase. In fact by (\ref{cached}) and (\ref{IVs}) the reducer node $k$ needs to compute $\{v_{q,n} : q \in [0,Q), w_n \notin B_{f}, B_f\in R_k,f \in [0,F) \}$, i.e., the set of IVs required for the output functions $\mathcal{W}_k$ from the files not accessible to it (from the files in $B_f$ such that $ f \in [0,F)$ and $p_{f,k}$ $\neq *$). Without loss of generality, let $p_{f,k} = s \in S_k$. For each $k_i \in \{k_1,k_2,\ldots,k_{g_s-1}\}$ in (\ref{partition symbols}), it can compute the symbol $\U_{\mathcal{W}_k,B_f}^{k_i}$ from the coded symbol $X_s^{k_i}$ transmitted by the reducer node $k_i$, i.e., 
\begin{align}
\label{node l transmission}
X_s^{k_i} = \bigoplus_{(u,v)\in [0,F) \times ([0,K) \backslash k_i) : p_{u,v}=s} \U_{\mathcal{W}_v,B_u}^{{k_i}}.
\end{align}
In (\ref{node l transmission}), for $v$ $ \neq k$, $p_{u,v} = p_{f,k}=s$ implies that $p_{u,k} = *$ by C2-2. Hence, the reducer node $k$ can compute $\U_{\mathcal{W}_v,B_u}^{k_i}$ by (\ref{cached}), (\ref{IVs}), (\ref{symbols_pda}) and (\ref{partition symbols}). For $v = k$, $p_{u,v} = p_{f,k} = s$ implies $u = f$ by C2-1. Therefore, the reducer node $k$ can compute the symbol $\U_{\mathcal{W}_k,B_f}^{k_i}$ from the coded symbol in (\ref{node l transmission}) by canceling out the rest of the symbols. By collecting all the symbols $\U_{\mathcal{W}_k,B_f}^{k_i}$ in (\ref{partition symbols}), the reducer node $k$ can compute the output functions in $\mathcal{W}_k$.

Now, we compute the communication load for this scheme. 
For each $s \in [0,S)$ occurring $g_s$ times, there are $g_s$ associated sequences sent, each of size $\frac{\eta_1 \eta_2 t}{(g_s-1)}$ bits by (\ref{transmission}). Let $S_g $ denotes the number of integers which appears exactly  $g$ times in the array. 
The communication load is given by
\begin{align}
L &= \frac{1}{QNt}\sum_{s=0}^{S-1} \frac{ g_s \eta_1 \eta_2 t}{(g_s-1)} 
\nonumber\\&= \frac{\eta_1 \eta_2 t}{\eta_1 \eta_2 KFt }\sum_{g=2}^{K} \frac{gS_g}{(g-1)} \nonumber\\
&=  \frac{\sum_{g=2}^{K} S_g}{KF}+\sum_{g=2}^{K} \frac{S_g}{KF (g-1)} 
\nonumber\\&=  \frac{S}{KF}+\sum_{g=2}^{K} \frac{S_g}{KF (g-1)}. 
\end{align}
\section{Proof of Theorem \ref{thm2}}
\label{proof thm2}
We present the proof of Theorem \ref{thm2} in this section. Based on an $(K, F, S)$ MRA $P = [p_{f,k}]$ with $f \in [0, F)$
and $k \in [0, K)$, an MADC scheme for an MRG having $F$ batches, $F$ mapper nodes and $K$ reducer nodes can be obtained as follows. We consider $Q=\eta_2 K$ output functions, for some integer $\eta_2$, so that each reducer node is assigned $\eta_2$ output functions to compute.

First, files are divided by grouping $N$ files into $F$ disjoint batches $\{B_{0},B_{1},\ldots, B_{F-1}\}$ each containing $ \eta_1 =  N / F$ files such that $\bigcup_{m=0}^{F-1} B_{m} = \{w_0,w_1,\ldots,w_{N-1}\}$. The rows in the MRA represent these $F$ batches $\{B_{0},B_{1},\ldots, B_{F-1}\}$. The columns in the MRA represent the  $K$ reducer nodes. Each mapper node $f \in [0,F)$ is assigned a batch $M_{f} = \{B_f\}$. Hence, the computation load is $r=1$. 
For every $f \in [0,F)$, the  mapper node $f$ computes IVs in the set $\{v_{q,n} : q \in [0,Q), w_n \in B_{f}\}$,
where each $v_{q,n}$ is a bit stream of length $t$. For each $k \in [0,K)$, the reducer node $k$ is connected to the mapper node $f$, if $p_{f,k} = *$, for $f \in [0,F)$. Each reducer node $k \in [0,K)$ can access all the batches in the set $
R_k = \{B_{f} : p_{f,k} = *, f \in [0,F)\}$. Hence, the shuffle and reduce phases follow from the proof of Theorem \ref{thm1} and using Theorem \ref{thm1}, the achievable communication load is given by  $L(1)= \frac{S}{KF}+\sum_{g=2}^{K}\frac{S_g}{KF (g-1)}$,  where $S_g$ is the number of integers in $[0,S)$ which appears exactly $g$ times in the MRA $P$. 
\section{MRAs for MADC Models with NNC-MRG}
\label{proof thm3}
In this section, we present the proof of Theorem \ref{thm3}.
In the map phase, we split the $N$ files into $\Lambda$ batches, $\{B_0,B_1,\ldots ,B_{\Lambda-1}\}$. Each mapper node $\lambda \in [0,\Lambda)$ is filled with batches of files as follows:
\begin{align}
M_{\lambda} = \{B_{(r\lambda+j) \text{ mod }\Lambda}: j \in [0,r)\}
\end{align}
Each mapper node stores $r$ batches of files. 
Each reducer node can access $\alpha$ mapper nodes and each mapper node has  $r$ consecutive batches of files. 
Hence, each reducer node has access to $\alpha r$ consecutive batches of files since the content in any consecutive $\alpha $ mapper nodes are disjoint from one another. That is, for each reducer node $\lambda \in [0,\Lambda)$, the set of all batches accessible to it is $\{B_{(r\lambda+j) \text{ mod } \Lambda}: j \in [0,r\alpha)\}$. 

Now, consider the $r$-cyclic $\frac{2\Lambda}{\Lambda-r(\alpha -1)}$-regular $\left( \Lambda,\Lambda,\alpha r, \frac{(\Lambda-\alpha r)(\Lambda-r(\alpha-1))}{2}\right)$ PDA constructed using Algorithm 2 in \cite{SR}. For this PDA, each integer appears exactly $\frac{2\Lambda}{\Lambda-r(\alpha -1)}$ times, which is greater than 1. Hence it is an $\frac{2\Lambda}{\Lambda-r(\alpha -1)}$-regular $\left( \Lambda,\Lambda, \frac{(\Lambda-\alpha r)(\Lambda-r(\alpha-1))}{2}\right)$ MRA. The rows of this MRA represent the batches and columns represent the reducer nodes. In this MRA, there are $\alpha r$ consecutive $*$ in the first column starting from the first row, which implies the reducer node $0$ has access to first $\alpha r$ batches of files. All other columns are obtained by shifting the previous column down by $r$ units. This matches our configuration. Hence, this MRA represents the NNC-MRG. Therefore, the shuffle and reduce phases follow from the proof of Theorem \ref{thm1} and using Theorem \ref{thm1}, the achievable communication load is given by  $L(r)= \frac{S}{KF}+\sum_{g=2}^{K}\frac{S_g}{KF (g-1)}$,  where $S_g$ is the number of integers in $[0,S)$ which appears exactly $g$ times in the MRA. In this case all the integers appear exactly $\frac{2\Lambda}{\Lambda-(\alpha -1)r}$ times. Hence, we have
\begin{align}
L(r) &= \frac{S}{KF}+\frac{S}{KF \left(\frac{2\Lambda}{\Lambda-(\alpha -1)r}-1\right)} \nonumber\\&= \frac{(\Lambda-\alpha r)(\Lambda-(\alpha-1)r) (2\Lambda)}{2\Lambda^2(\Lambda+(\alpha-1)r)} \nonumber\\&=\frac{(\Lambda-\alpha r)(\Lambda-(\alpha-1)r)}{\Lambda (\Lambda+(\alpha-1)r)}.
\end{align}

\section{Proof of correctness of {\textbf{Algorithm \ref{algo1}}}}
\label{proof algo1}
In this section, we present the proof of correctness of {\textbf{Algorithm \ref{algo1}}}, i.e., we prove that the array obtained using  {\textbf{Algorithm \ref{algo1}}} corresponds to a $g-$regular MRA.

In {\textbf{procedure 1}}, all subsets of size $\alpha +r$ from $[0,\Lambda)$ are arranged in lexicographic order and for any subset $T'$ of size $\alpha+r$, we define $y_{\alpha+r}(T')$ to be its order minus 1. Clearly, $y_{\alpha+r}$ is a bijection from $\{T' \subset [0,\Lambda) : |T'| = \alpha+r \}$ to $\left[0, {\Lambda \choose \alpha+r} \right)$. For example, when $\Lambda = 5,\alpha = 2$ and $r=2$, all the subsets of size $\alpha +r=4$ in $\{0, 1, 2, 3,4\}$ are ordered as $\{0, 1,2,3\},\{0, 1,2,4\},\{0, 1,3,4\},\{0,2,3,4\}$ and $\{1,2, 3,4\}.$ Accordingly, $
y_4{(0, 1,2,3)} = 0, y_4{(0, 1,2,4)} = 1, y_4{(0, 1,3,4)} = 2,y_4{(0,2,3,4)} = 3, $ and $y_4{(1,2, 3,4)} = 4.$

In {\textbf{procedure 2}}, we define an ${\Lambda \choose r} \times {\Lambda \choose \alpha}$ array $D_{\Lambda,r,\alpha}$. The rows of which are denoted by the sets in $\{ T\subset [0,\Lambda), |T| = r \}$ and columns by the sets in $\{U \subset [0,\Lambda) : |U| = \alpha\}$. 
Each entry $d_{T,U}$ corresponding to the row $T$ and the column $U$ are obtained as (\ref{Dk}).
We next prove that $D_{\Lambda,r,\alpha}$ is a $g-$regular MRA. 
From (\ref{Dk}), $D_{\Lambda,r,\alpha}$ is an ${\Lambda \choose r} \times {\Lambda \choose \alpha}$ array consisting of $*$ and integers in $\left[0,{\Lambda \choose  \alpha +r}\right)$. Hence, $S = {\Lambda \choose  \alpha +r}$. Next, we need to check if it obeys $ C1'$, and $C2$.
From (\ref{Dk}), it is clear that the symbol $*$ appears if $|T \cap U |\neq 0,$ i.e., if $T$ and $U$ have some integer in common. For a given $U$, since $|U| =\alpha,$ there are ${\Lambda -\alpha \choose r}$ ways in which we can select $T$ such that $|T \cap U| = 0$. Hence, the symbol $*$ appears exactly $ {\Lambda \choose r} - {\Lambda -\alpha \choose r} $ times in each column $U$.
Next, consider two distinct entries $d_{{T_1},\U_1} = d_{{T_2},\U_2} = s$, where $T_1, T_2, \U_1, \U_2 \subset [0,\Lambda)$ with $|T_1|=|T_2| = r$ and $|\U_1|=|\U_2| = \alpha$. Applying the fact that $y_{\alpha+r}$ is a bijection from $\{T' \subset [0,\Lambda) : |T'| = \alpha + r\}$ to $\left[0, {\Lambda \choose \alpha +r}\right)$ from (\ref{Dk}), we know that $s$ is an integer if and only if $T_1 \cup \U_1 = T_2 \cup \U_2,$ which implies that
\begin{itemize}
	\item Each integer $y_{\alpha+r}(T')$ in $\left[0, {\Lambda \choose \alpha +r}\right)$ occurs exactly ${r+\alpha \choose r}$ times since for a given $T'$, there are ${r+\alpha \choose r}$ distinct possibilities of $(\{T: T \subset T', |T| = r\}, U = P \backslash T)$ (since $|T'| = \alpha + r$). Thus $C1'$ is satisfied.
	\item  $T_1 \neq T_2$ and $\U_1 \neq \U_2$, i.e. the two entries are in distinct rows and columns. Further, this is equivalent to $|\U_1 \cap T_2| \neq 0$ and $ |\U_2 \cap T_1| \neq 0,$ (since $T_1 \cup \U_1 = T_2 \cup \U_2$). Thus, $d_{{T_1},\U_2} = d_{{T_2},\U_1} = *$ by (\ref{Dk}), and, hence, $C2$ is satisfied.
\end{itemize}
In other words, both of the conditions $C1'$, and $C2$ hold. That concludes the proof of correctness of {\textbf{Algorithm \ref{algo1}}}.

\section{Representation of  MADC Scheme in \cite{BP} with CT via MRAs}
\label{proof thm4}
In this section, we prove Theorem \ref{thm4}, i.e., we prove that the $g-$regular MRA $D_{\Lambda,r,\alpha}$ obtained using {\textbf{Algorithm \ref{algo1}}} represents MADC models with CT with $\Lambda$ mapper nodes, and $K$ reducer nodes.

Recall that the input database is split into $F={\Lambda \choose r}$ disjoint batches $B_{T}$ with $ T \subset [0,\Lambda)$ and $ |T|=r$. The mapper node $\lambda \in [0,\Lambda)$ is assigned all batches $B_{T}$ if $\lambda \in T$.
We have $K = {\Lambda \choose \alpha}$ reducer nodes, where there is a unique reducer node connected to each subset of $\alpha$ mapper nodes and each reducer node is labeled by a subset of size $\alpha$ in the set $\{0, 1,\ldots , \Lambda-1\}$.
It can be observed that the array $D_{\Lambda,r,\alpha}$ corresponds to an MADC model with CT, with the rows corresponding to the batches and the column corresponding to the reducer nodes. There is a $*$ in an entry  corresponding to the row $T$ and column $U$ if and only if the reducer node has access to the batch $B_T$, i.e., if and only if $|T\cap U| \neq 0$. This matches our model.
The shuffle and reduce phases follow from the proof of Theorem \ref{thm1}. Hence, the communication load is given by 
$     L = \frac{S}{KF}+\sum_{g=2}^{K}\frac{S_g}{KF (g-1)} ,$
where $S_g$ is the number of integers in $[0,S)$ which appears exactly $g$ times in the MRA $D_{\Lambda,r,\alpha}$. In this case all the integers appear exactly ${r+\alpha \choose r}$ times. Hence, we have
\begin{align}
L(r) &= \frac{S}{KF}+\frac{S}{KF \left({r+\alpha \choose r}-1\right)} \nonumber\\
&= \frac{{\Lambda \choose \alpha + r}}{{\Lambda \choose \alpha}{\Lambda \choose r} } \left (1+ \frac{1}{{r+\alpha \choose r}-1}\right ) \nonumber\\
&= \frac{{\Lambda \choose \alpha + r}{r+\alpha \choose r}}{{\Lambda \choose \alpha}{\Lambda \choose r} \left({r+\alpha \choose r}-1\right)} \nonumber\\
&= \frac{\Lambda !}{(\alpha+r)! (\Lambda-\alpha-r)!} \times \frac{(\Lambda-\alpha)! \alpha !}{\Lambda !}\nonumber\\ &\hspace{0.5cm}\times \frac{(\alpha +r) ! }{\alpha ! r!} \times\frac{ 1}{{\Lambda \choose r} \left({r+\alpha \choose r}-1\right)} \nonumber\\
&=  \frac{(\Lambda-\alpha)! }{(\Lambda-\alpha-r)! r!} \times \frac{ 1}{{\Lambda \choose r} \left({r+\alpha \choose r}-1\right)} \nonumber\\
&= \frac{ {\Lambda-\alpha \choose r }}{{\Lambda \choose r} \left({r+\alpha \choose r}-1\right)}. 
\end{align}

\section{Proof of correctness of \textbf{Algorithm \ref{algo2}}}
\label{proof algo2}
Using \textbf{procedure 1} of \textbf{Algorithm \ref{algo2}}, we construct a  ${r+\alpha \choose r}\text{-regular }\left ({\Lambda \choose \alpha}, {\Lambda \choose r},  {\Lambda \choose  \alpha +r} \right ) $ MRA $D_{\alpha,1}$ for each $\alpha \in [\Lambda-r]$.

Using the \textbf{procedure 2},  for each $\alpha \in [\Lambda-r]$, we define new arrays ${ D}_{\alpha,m},$ for $ m \in [2,K_{\alpha}]$, where each entry in $D_{\alpha,m}$ is obtained by adding $(m-1)S_1$ to the corresponding entry in $D_{\alpha,1}$. ${S}_1$ denotes the number of integers present in the array $D_{\alpha,1}$, and $ *+(m-1){S}_1  =*$.  The array ${\bf D}_{\alpha}$ is obtained by concatenating the $K_{\alpha}$ arrays $D_{\alpha,m},$ for $ m \in [K_{\alpha}]$, each of size ${\Lambda \choose r} \times {\Lambda \choose \alpha}$. Hence the array ${\bf D}_{\alpha}$ is of size ${\Lambda \choose r} \times K_{\alpha}{\Lambda \choose \alpha}.$

Each array ${ D}_{\alpha,m},$ for $ m \in [2,K_{\alpha}]$, represents a ${r+\alpha \choose r}\text{-regular }\left ({\Lambda \choose \alpha}, {\Lambda \choose r}, {\Lambda \choose  \alpha +r} \right ) $ MRA with the ${\Lambda \choose  \alpha +r}$ integers present in the MRA ${ D}_{\alpha,m}$ being $[(m-1){S}_1,m{S}_1)$. 	This is because, each entry in $ { D}_{\alpha,m}$ is obtained by adding $(m-1){S}_1$ to the corresponding entry in ${ D}_{\alpha,1}$, where $* +(m-1){S}_1 =*.$

The array ${\bf D}_{\alpha}$ is obtained by concatenating $K_{\alpha}$ number of ${r+\alpha \choose r}\text{-regular }\left ({\Lambda \choose \alpha}, {\Lambda \choose r},  {\Lambda \choose  \alpha +r} \right ) $ MRAs with the integers present in each of the MRAs  ${ D}_{\alpha,m}$, for $ m\in [K_{\alpha}],$ are different from one another.
Hence, the total number of integers present in the array ${\bf D}_{\alpha}$ obtained by concatenating all the arrays in $\{{ D}_{\alpha,m}: m \in [K_{\alpha}]\}$ is $S=K_{\alpha }{S}_1=K_{\alpha} {\Lambda \choose \alpha +r}$ and the array ${\bf D}_{\alpha}$ obeys condition $\textit{C1}'$ in Definition \ref{def:g-MRA} with $g={r+\alpha \choose r}$.  The array ${\bf D}_{\alpha}$ also satisfies condition $\textit{C2}$ in Definition \ref{def:GPDA}. Hence the array ${\bf D}_{\alpha}$ represents a  ${r+\alpha \choose r}\text{-regular }\left ( K_{\alpha}{\Lambda \choose \alpha}, {\Lambda \choose r}, K_{\alpha}{\Lambda \choose  \alpha +r} \right ) $ MRA. 

Using \textbf{procedure 3}, we define new arrays $\tilde{\textbf{D}}_{\alpha}$ for $ \alpha \in [2,\Lambda-r]$, where each entry in $\tilde{\textbf{D}}_{\alpha}$ is obtained by adding $S'_{\alpha}$ to the corresponding entry in ${\textbf{D}}_{\alpha}$, with $*+S'_{\alpha}  =*$.
For every  $\alpha \in [2,\Lambda -r]$, $S'_{\alpha}$ denotes the total number of integers present in the set of arrays $\{\tilde{\textbf{D}}_{m}:m\in [\alpha-1]\}$. The array ${\bf D}$ is obtained by concatenating the $\Lambda-r$ arrays $\tilde{\textbf{D}}_{\alpha},$ for $ \alpha\in [\Lambda-r]$, each of size ${\Lambda \choose r} \times K_{\alpha}{\Lambda \choose \alpha}$. Hence the array ${\bf D}$ is of size ${\Lambda \choose r} \times \sum_{\alpha\in [\Lambda-r]}K_{\alpha}{\Lambda \choose \alpha}.$

Each array $\tilde{ \bf D}_{\alpha},$ for $ \alpha \in [2,\Lambda-r]$ represents a ${r+\alpha \choose r}\text{-regular }\left (K_{\alpha}{\Lambda \choose \alpha}, K_{\alpha}{\Lambda \choose  \alpha +r} \right ) $ MRA with the $K_{\alpha}{\Lambda \choose  \alpha +r}$ integers present in the MRA $\tilde{ \bf D}_{\alpha}$ being $[S'_{\alpha},S'_{\alpha}+K_{\alpha}{\Lambda \choose  \alpha +r})$. 	This is because, each entry in $ \tilde{ \bf D}_{\alpha}$ is obtained by adding $S'_{\alpha}$ to the corresponding entry in ${ \bf D}_{\alpha}$, where $* +S'_{\alpha} =*.$

The array ${\bf D}$ is obtained by concatenating $\Lambda-r$ number of ${r+\alpha \choose r}\text{-regular }\left (K_{\alpha}{\Lambda \choose \alpha}, {\Lambda \choose r}, K_{\alpha} {\Lambda \choose  \alpha +r} \right ) $ MRAs with the integers present in each of the MRAs  $\tilde{ \bf D}_{\alpha}$, for $ \alpha\in [\Lambda-r],$ are different from one another.
The total number of integers present in the array ${\bf D}$ obtained by concatenating all the arrays is $S=\sum_{\alpha\in [\Lambda-r]}K_{\alpha } {\Lambda \choose \alpha +r}$ and the array ${\bf D}$ obeys conditions $\textit{C1}$ and $\textit{C2}$ in Definition \ref{def:GPDA}. Hence, the array ${\bf D}$ is a $\left ( \sum_{\alpha \in [\Lambda-r]}K_{\alpha} {\Lambda \choose \alpha} , {\Lambda \choose r}, \sum_{\alpha\in [\Lambda-r]} K_{\alpha}{\Lambda \choose  \alpha +r} \right ) $ MRA. This completes the proof.

\section{MRAs for MADC Models with GC-MRG}
\label{proof thm5}
In this section, we prove Theorem \ref{thm5}, i.e., we prove that the MRA ${\bf D}$ obtained using {\textbf{Algorithm \ref{algo2}}} represents MADC models with GC-MRG with $\Lambda$ mapper nodes and $K$ reducer nodes.

Recall that similar to MADC models with CT, the input database is split into $F={\Lambda \choose r}$ disjoint batches $B_{T}$ with $ T \subset [0,\Lambda)$ and $ |T|=r$. The mapper node $\lambda \in [0,\Lambda)$ is assigned a batch $B_{T}$ if $\lambda \in T$.
We have $K = \sum_{\alpha \in [\alpha-r]}K_{\alpha}{\Lambda \choose \alpha}$ reducer nodes, where for every combination of $\alpha$ mapper nodes, $\alpha\in [\Lambda-r]$,  there are $K_{\alpha}$ reducer nodes to which those mapper nodes are uniquely connected to.

Like we described in Section \ref{GC-MRG}, we divide $K$ users into $\Lambda-r$ disjoint blocks, $\{A_{\alpha} :\alpha \in [\Lambda-r]\}$, such that all the reducer nodes connected to exactly $\alpha$ mapper nodes are put in the block $A_{\alpha}$, for each $\alpha \in [\Lambda-r]$. Each block $A_{\alpha}$ contains $K_{\alpha}{\Lambda \choose \alpha}$ reducer nodes. Furthermore, we sub-divide each  block $A_{\alpha}$, for $\alpha \in [\Lambda-r]$, into $K_{\alpha}$ disjoint sub-blocks $\{A_{\alpha,m}:m \in [K_{\alpha}]\}$ such that each sub-block $A_{\alpha,m}$ contains a set of reducer nodes where no two reducer nodes are  connected to the same set of $\alpha$ mapper nodes. Hence, each sub-block $A_{\alpha,m}$ contains ${\Lambda \choose \alpha}$ reducer nodes and they represent MADC models with CT, for $m \in [K_{\alpha}]$.

We construct an  MRA {\bf D} using {\bf Algorithm \ref{algo2}}. The rows correspond to the batches of files.
We observe that for each $\alpha \in [\Lambda-r]$, and $m \in [K_{\alpha}]$, the columns in the array ${ D}_{\alpha,m}$ correspond to the reducer nodes in the set $A_{\alpha,m}$. Hence, the columns in the array ${\bf D}$ correspond to the reducer nodes in the set $\cup_{\alpha \in [\Lambda-r],m \in [K_{\alpha}]} A_{\alpha,m}$ which is the set of reducer nodes we are interested in. 
There is a $*$ in an entry  corresponding to the row $T$ and column $k$ if and only if the reducer node corresponding to the column $k\in [0,K)$ has access to the batch $B_T$.  So, the array ${\bf D}$ corresponds to an MADC model with GC-MRG, with the rows corresponding to the batches and the column corresponding to the reducer nodes. 
Hence, the shuffling and reduce phases follow from the proof of Theorem \ref{thm1} and the communication load is given by 
$     L = \frac{S}{KF}+\sum_{g=2}^{K}\frac{S_g}{KF (g-1)} ,$
where $S_g$ is the number of integers in $[0,S)$ which appears exactly $g$ times in the MRA ${\bf D}$. In this case all the integers in the range $[S'_{\alpha},S'_{\alpha}+K_{\alpha}{\Lambda \choose  \alpha +r})$ appear exactly ${r+\alpha \choose r}$ times, for each $\alpha \in [\Lambda-r]$. Hence, $K_{\alpha}{\Lambda \choose  \alpha +r}$ integers appear exactly ${r+\alpha \choose r}$ times, for each $\alpha \in [\Lambda-r]$ and we have
\begin{align}
L &= \sum_{\alpha\in [\Lambda-r]} \frac{K_{\alpha}{\Lambda \choose  \alpha +r}}{KF}+\sum_{\alpha\in[\Lambda-r]}\frac{K_{\alpha}{\Lambda \choose  \alpha +r}}{KF \left({r+\alpha \choose r}-1\right)} \nonumber\\
&= \sum_{\alpha\in [\Lambda-r]} \frac{K_{\alpha}{\Lambda \choose  \alpha +r}{r+\alpha \choose r}}{KF \left({r+\alpha \choose r}-1\right)} \nonumber\\
&= \sum_{\alpha\in [\Lambda-r]} \frac{K_{\alpha}{\Lambda \choose \alpha + r}{r+\alpha \choose r}}{K{\Lambda \choose r} \left({r+\alpha \choose r}-1\right)} \nonumber\\
&= \frac{1}{K} \sum_{\alpha\in [\Lambda-r]} \frac{K_\alpha \Lambda !}{(\alpha+r)! (\Lambda-\alpha-r)!} \times \frac{(\Lambda-r)! r !}{\Lambda !}\nonumber\\ &\hspace{1.5cm}\times \frac{(\alpha +r) ! }{\alpha ! r!} \times\frac{ 1}{ \left({r+\alpha \choose r}-1\right)} \nonumber\\
&= \frac{1}{\left (\sum_{\alpha \in [\Lambda-r]} K_{\alpha} {\Lambda \choose \alpha} \right )}\sum_{\alpha\in [\Lambda-r]}\frac{ K_{\alpha}{\Lambda-r \choose \alpha }}{ \left({r+\alpha \choose r}-1\right)}. 
\end{align}
\section{Lower Bound for Homogeneous Networks with GC-MRG}
\label{proof thm6}
In this section we provide a lower bound for MADC models operating on homogeneous network with GC-MRG.

Recall that $M_{\lambda} $ denotes the set of all files mapped by the mapper node $\lambda \in [0,\Lambda)$ and $R_k$ denotes the set of files accessible to the reducer node $k \in [0,K)$. The files are divided by grouping $N$ files into $F$ disjoint batches each containing $ \eta_1 =  N / F$ files. Each reducer node is assigned $\eta_2 =\frac{Q}{K}$ output functions to compute.

Consider a file assignment $\M =\{\M_0,\M_1,\ldots,\M_{\Lambda-1}\}$ in the Map phase, where $\M_{\lambda}$, for $ \lambda \in [0,\Lambda)$, denotes the set of all files assigned to the mapper node $\lambda$. Consider the  file assignment at the reducer nodes to be $\R =\{\R_0,\R_1,\ldots,\R_{K-1}\}$, where $\R_{k}$, for $k \in [0,K) $, denotes the set of all files accessible to the reducer node $k$. For this mapper-reducer file assignment pair $(\M,\R)$, let the minimum communication load required be represented by $L_{\M, \R}^*$. 

We denote the number of files that are exclusively accessible to $j$ reducer nodes under this file assignment $(\M,\R)$ as $a_{\M,\R}^j$, for all $j \in [K]$, i.e., we have
\begin{align}
a_{\M,\R}^j  = \sum_{J \subseteq [K], |J|=j}|(\cap_{k \in J}\R_k)\backslash(\cup_{i \notin J} \R_i)|.
\end{align} 
For this file assignment, a lower bound on $L_{\M,\R}^*$ is given by 
\begin{align}
\label{lb for file assign}
L_{\M,\R}^* \geq \sum_{j = 1}^{K} \frac{a_{\M,\R}^j}{N}\frac{K-j}{Kj}.
\end{align}
The proof of (\ref{lb for file assign}) is similar to the proof of Lemma 1 in \cite{LMA}. For completeness we provide the proof in Appendix I.

The optimal communication load $L^*(r)$ is lower bounded by the minimum value of $L_{\M,\R}^*$ over all the possible file assignments which has a computation load of $r$. Hence,
\begin{align}
\label{convex}
L^*(r) &\geq \inf_{\substack{\text{$\M,\R:$} \\ \text{$|\M_0|+|\M_1|+$} \\ \text{$\ldots+|\M_{\lambda}| =rN$}}} L_{\M,\R}^* \nonumber\\
&\geq \inf_{\substack{\text{$\M,\R:$} \\ \text{$|\M_0|+|\M_1|+$} \\ \text{$\ldots+|\M_{\lambda}| =rN$}}}\sum_{j = 1}^{K} \frac{a_{\M,\R}^j}{N} \frac{K-j}{Kj}.
\end{align} 
Consider a file assignment $\M$ such that $|\M_0|+|\M_1|+\ldots+|\M_{\lambda}| =rN$. We have $a_{\M,\R}^j \geq 0, \forall j \in [K]$ and $\sum_{j \in [K]}a_{\M,\R}^j =N.$

For GC-MRG, for each file $n$, the number of reducer nodes having access to file $n$ is $\sum_{\alpha \in [\Lambda-r]}K_{\alpha}\left ({\Lambda \choose \alpha} - {\Lambda-r \choose \alpha} \right ) $. The sum of number of files accessible to the $K$ reducer nodes is
\begin{align}
\sum_{j \in [K]}ja_{\M,\R}^j &=\sum_{\alpha \in [\Lambda-r]}K_{\alpha}N\left ({\Lambda \choose \alpha} - {\Lambda-r \choose \alpha} \right ).
\end{align}
The function $\frac{K-j}{j}$ in (\ref{convex}) is convex in $j$, hence,
\begin{align}
L^*(r) &\geq \inf_{\substack{\text{$\M,\R:$} \\ \text{$|\M_0|+|\M_1|+$} \\ \text{$\ldots+|\M_{\lambda}| =rN$}}}
\frac{K\sum_{j = 1}^{K} \frac{a_{\M,\R}^j}{N}-\sum_{j = 1}^{K} j \frac{a_{\M,\R}^j}{N}}{K\sum_{j = 1}^{K} j \frac{a_{\M,\R}^j}{N}}\nonumber\\
&=  \frac{K-\sum_{\alpha \in [\Lambda-r]}K_{\alpha}\left ({\Lambda \choose \alpha} - {\Lambda-r \choose \alpha} \right ) }{K\sum_{\alpha \in [\Lambda-r]}K_{\alpha}\left ({\Lambda \choose \alpha} - {\Lambda-r \choose \alpha} \right ) }\nonumber\\
&=  \frac{\sum_{\alpha \in [\Lambda-r]}K_{\alpha}{\Lambda \choose \alpha}-\sum_{\alpha \in [\Lambda-r]}K_{\alpha}\left ({\Lambda \choose \alpha} - {\Lambda-r \choose \alpha} \right ) }{K\sum_{\alpha \in [\Lambda-r]}K_{\alpha}\left ({\Lambda \choose \alpha} - {\Lambda-r \choose \alpha} \right ) }\nonumber\\
&=  \frac{\sum_{\alpha \in [\Lambda-r]}K_{\alpha} {\Lambda-r \choose \alpha}  }{\left (\sum_{\alpha \in [\Lambda-r]} K_{\alpha} {\Lambda \choose \alpha} \right ) \left (\sum_{\alpha \in [\Lambda-r]}K_{\alpha}\left ({\Lambda \choose \alpha} - {\Lambda-r \choose \alpha} \right )\right ) }
\label{lb}
\end{align}
The lower bound on $L^*(r)$ in (\ref{lb}) holds for any non integer valued $r$ such that $1 \leq r \leq [\Lambda -1]$. The proof of which is similar to the proof provided in Section VI in \cite{LMA}. Hence, $L^*(r)$ is lower bounded by the lower convex envelope of the points in the set 
$\left \{\left ( r, \frac{\sum_{\alpha \in [\Lambda-r]}K_{\alpha} {\Lambda-r \choose \alpha}  }{K \left (\sum_{\alpha \in [\Lambda-r]}K_{\alpha}\left ({\Lambda \choose \alpha} - {\Lambda-r \choose \alpha} \right )\right ) }\right ): r \in [\Lambda -1] \right \}$.
\section{Proof related to Appendix \ref{proof thm6}}
We need to prove Eqn. (\ref{lb for file assign}). For $q \in [0,Q)$ and $n \in [0,N)$, let $V_{q,n}$ be i.i.d random variables uniformly distributed on $\F_{2^t}$ and the IVs $v_{q,n}$ be the realizations of $V_{q,n}$. For any subset $\Q \subseteq[0,Q), \N \subseteq \{w_0,w_1,\ldots,w_{n-1}\},$ we define $V_{\Q,\N} = \{V_{q,n}:q \in \Q, w_n\in \N\}$.

Recall that each reducer node $k \in [0,K)$ generates a coded symbol ${\bf X}_k$ using the IVs accessible to them (each reducer node $k$ has access to the IVs computed from the files in $\R_k$), hence $H({\bf X}_k|V_{:,\R_k}) =0,$ where $``:"$ denotes set of all possible indices. For any MADC scheme, each reducer node has to be able to recover all the IVs corresponding to the output functions which they need to compute. Thus, $H(V_{W_k,:} |{\bf X}_:,V_{:,\R_k}) = 0.$   For any subset $\mathcal{S} \subseteq [0,K)$, we define $\Y_{\mathcal{S}} = (V_{W_{\mathcal{S}}, :}, V_{:,\R_{\mathcal{S}}})$, which contains all the IVs  required by the nodes in $\mathcal{S}$ and all the IVs accessible to the nodes in $\mathcal{S}$. For some file assignment pair $(\M,\R)$ we denote the number of files that are exclusively accessible to $j$ nodes in $\mathcal{S}$ by $ a_{\M,\R}^{j,\mathcal{S}}  = \sum_{J \subseteq \mathcal{S}, |J|=j}|(\cap_{k \in J}\R_k)\backslash(\cup_{i \notin J} \R_i)|$ and denote the coded symbols sent by the nodes in $\mathcal{S}$ by $X_{\mathcal{S}} = \{X_k:k\in\mathcal{S}\}$.
For any $\mathcal{S} \subseteq [0,K)$, we have $H(X_{\mathcal{S}}|\Y_{\mathcal{S}^c}) \geq t\sum_{j=1}^{|\mathcal{S}|}a_{\M,\R}^{j,\mathcal{S}} \frac{Q}{K}\frac{|\mathcal{S}|-j}{j}$, where $\mathcal{S}^c =[0,K) \backslash \mathcal{S}$. The proof of which is similar to the one provided in  Lemma 1 \cite{LMA} where this is proved by induction.
Let $\mathcal{S} =[0,K)$ be the set of all $K$ reducer nodes. Then, 
\begin{align}
L_{\M,\R}^{*} \geq \frac{H(X_{\mathcal{S}} | \Y_{\mathcal{S}^c})}{QNt} \geq \sum_{j = 1}^{K} \frac{a_{\M,\R}^j}{N}\frac{K-j}{Kj}.
\end{align}
\end{document}